\pdfoutput=1
\documentclass[runningheads]{llncs}
\usepackage{graphicx}

\newif\ifcav
\cavfalse

\usepackage{amsmath}
\usepackage{amsfonts}
\usepackage{amssymb}
\usepackage{graphicx}

\usepackage{hyperref}
\usepackage{cleveref}

\usepackage{url}

\usepackage{subcaption}
\usepackage{wrapfig}

\usepackage{algorithm}
\usepackage{algorithmicx}
\usepackage[noend]{algpseudocode}
\usepackage{tikz}
\usetikzlibrary{arrows, patterns, math, positioning}
\usepackage{pgfplots}

\usepackage{blkarray}
\usepackage{tabularx}
\usepackage{multirow}
\usepackage{enumitem}

\newcommand{\OO}{o}%{\mathbb{O}}
\newcommand{\II}{i}%{\mathbb{I}}
\newcommand{\W}{w}%{\mathbb{W}}
\newcommand{\V}{v}%{\mathbb{V}}

\newcommand{\true}{\top}
\newcommand{\false}{\bot}

\newcommand{\OR}{\vee}
\newcommand{\AND}{\wedge}

\renewcommand{\implies}{\Rightarrow}

\newcommand\equant[1]{{\exists #1}}
\newcommand\uquant[1]{{\forall #1}}

\newcommand\dom[1]{\mathcal{D}(#1)}
\newcommand{\sequiv}{\equiv}

\newcommand{\lequiv}{\Leftrightarrow}

\newcommand{\code}[1]{\text{\texttt{#1}}}

\newcommand{\modabs}{\preceq}
\newcommand{\comp}{\code{comp}}

\newcommand\quant[1]{Q_{#1}}
\newcommand\abs[1]{\hat{#1}}
\newcommand{\void}{\emptyset}

\newcommand{\CPRE}{{\code{cpre}}}

\newcommand{\ihide}{\code{ihide}}
\newcommand{\ohide}{\code{ohide}}

\newcommand{\incoarsen}{\code{icoarsen}}
\newcommand{\outcoarsen}{\code{ocoarsen}}
\newcommand{\refine}{\code{refine}}

\newcommand\NB[1]{\code{nb}(#1)}%{\textsf{NB}_{#1}}

\newcommand\pref[1]{{(\ref{#1})}}

\renewcommand{\emptyset}{{\varnothing}}

\begin{document}
%
%\ifcav
\title{Flexible Computational Pipelines for Robust Abstraction-Based Control Synthesis \thanks{The authors were funded in part by AFOSR FA9550-18-1-0253, DARPA Assured Autonomy project, iCyPhy, Berkeley Deep Drive, and NSF grant CNS-1739816.}
}
\titlerunning{Flexible Pipelines for Abstraction-Based Control Synthesis}
%\author{Authors Omitted for Review}
\author{Eric S. Kim \orcidID{0000-0002-2926-7994} \and
Murat Arcak \orcidID{0000-0001-9060-4032} \and%\inst{2,3} \and
Sanjit A. Seshia \orcidID{0000-0001-6190-8707}%\inst{3}
}
\institute{UC Berkeley, Berkeley, CA, USA\\ %\and
%%\url{http://www.springer.com/gp/computer-science/lncs} \and
%%ABC Institute, Rupert-Karls-University Heidelberg, Heidelberg, Germany\\
\email{\{eskim,arcak,sseshia\}@eecs.berkeley.edu}}
%
%\else
%\title{Flexible Computational Pipelines for Robust Abstraction-Based Control Synthesis \thanks{The authors were funded in part by AFOSR FA9550-18-1-0253, DARPA Assured Autonomy project, iCyPhy, Berkeley Deep Drive, and NSF grant CNS-1739816.}
%}
%\author{Eric S. Kim, Murat Arcak, Sanjit A. Seshia\\
%UC Berkeley, Berkeley, CA, USA\\
%\{eskim,arcak,sseshia\}@eecs.berkeley.edu}
%\date{}
%\newtheorem{definition}{Definition}
%\newtheorem{example}{Example}
%\newtheorem{proof}{Proof}
%\newtheorem{proposition}{Proposition}
%\newtheorem{theorem}{Theorem}
%\newtheorem{corollary}{corollary}
%\newtheorem{remark}{Remark}
%\fi

\maketitle              % typeset the header of the contribution
\begin{abstract}
Successfully synthesizing controllers for complex dynamical systems and specifications often requires leveraging domain knowledge as well as making difficult computational or mathematical tradeoffs. This paper presents a flexible and extensible framework for constructing robust control synthesis algorithms and applies this to the traditional abstraction-based control synthesis pipeline. It is grounded in the theory of relational interfaces and provides a principled methodology to seamlessly combine different techniques (such as dynamic precision grids, refining abstractions while synthesizing, or decomposed control predecessors) or create custom procedures to exploit an application's intrinsic structural properties. A Dubins vehicle is used as a motivating example to showcase memory and runtime improvements.
%This paper presents an extensible framework for control synthesis through finite abstraction. We use atomic operators for interface composition, refinement, and variable hiding from the theory of relational interfaces to create new composite operators that capture common procedures in existing tools like abstraction and controlled predecessors. It also provides a principled methodology to seamlessly and flexibly combine techniques such as dynamic precision grids and merging abstraction and synthesis algorithms. Users can integrate custom procedures to best exploit an application's intrinsic structural properties. A Dubins vehicle is used as a motivating example to showcase memory and runtime improvements.

\ifcav
\else
\fi
\end{abstract}

\section{Introduction}
A control synthesizer's high level goal is to automatically construct control software that enables a closed loop system to satisfy a desired specification. 
A vast and rich literature contains results that mathematically characterize solutions to different classes of problems and specifications, such as the Hamilton-Jacobi-Isaacs PDE for differential games \cite{basar1999dynamic}, Lyapunov theory for stabilization \cite{khalil2002nonlinear}, and fixed-points for temporal logic specifications \cite{piterman2006synthesis}\cite{majumdar2003symbolic}.
While many control synthesis problems have elegant mathematical solutions, there is often a gap between a solution's theoretical characterization and the algorithms used to compute it.
What data structures are used to represent the dynamics and constraints? What operations should those data structures support?
How should the control synthesis algorithm be structured?
Implementing solutions to the questions above can require substantial time.
This problem is especially critical for computationally challenging problems, where it is often necessary to let the user \textit{rapidly} identify and exploit structure through analysis or experimentation.

\subsection{Bottlenecks in Abstraction-based Control Synthesis} \label{subsec:abcs}
This paper's goal is to enable a framework to develop extensible tools for robust controller synthesis.
It was inspired in part by computational bottlenecks encountered in control synthesizers that construct finite abstractions of continuous systems, which we use as a target use case. A traditional abstraction-based control synthesis pipeline consists of three distinct stages:
\begin{enumerate}
\item Abstracting the continuous state system into a finite automaton whose underlying transitions faithfully mimic the original dynamics \cite{Tabu}, \cite{majid}.
\item Synthesizing a discrete controller by leveraging data structures and symbolic reasoning algorithms to mitigate combinatorial state explosion.
\item Refining the discrete controller into a continuous one. Feasibility of this step is ensured through the abstraction step.
\end{enumerate}
This pipeline appears in tools PESSOA \cite{mazo2010pessoa} and SCOTS \cite{Rungger2016}, which can exhibit acute computational bottlenecks for high dimensional and nonlinear system dynamics.
%For instance, constructing finite abstractions requires an enumerative traversal of the state-input space that grows exponentially with dimension.
%Even though these tools are designed to automatically output a ``correct-by-construction" controller, their implementations are designed for generic systems and do not exploit structure.
%The abstraction step's and utility of the outputted controller can be highly sensitive to abstraction parameters.
%These parameters are typically chosen in a pre-processing step but the best parameters are often only known retroactively.
% and partitioning is a preprocessing step, which is critical in determining the computational feasibility of the abstraction and synthesis steps. 
A common method to mitigate these bottlenecks is to exploit a specific dynamical system's topological and algebraic properties. In MASCOT \cite{Hsu:2018:MAC:3178126.3178143} and CoSyMA \cite{mouelhi2013cosyma}, multi-scale grids and hierarchical models capture notions of state-space locality. One could incrementally construct an abstraction of the system dynamics while performing the control synthesis step \cite{nilsson2017augmented} \cite{Liu:2017:RAC:3049797.3049826} as implemented in tools ROCS \cite{rocs} and ARCS \cite{novelbddencoding}.
The abstraction overhead can also be reduced by representing systems as a collection of components composed in parallel \cite{meyer} \cite{sparseabs2017}. These have been developed in isolation and were not previously interoperable.
%\todo{Control related work, abs int related work, relational interfaces related work.}

%Control synthesis tools implementing strict subsets of the solutions above include PESSOA \cite{mazo2010pessoa}, TuLiP \cite{wongpiromsarn2011tulip} ROCS \cite{rocs}, SCOTS \cite{Rungger2016}, ARCS \cite{novelbddencoding}, CoSyMa \cite{mouelhi2013cosyma}, and MASCOT \cite{Hsu:2018:MAC:3178126.3178143}, \todo{SLUGS?}.
%Tackling higher dimensional problems requires a combination of these techniques, but merging these tools is a non-trivial task because the tools differ in fundamental notions such as the source of non-determinism or whether grids are a partition with hyper rectangles or a collection of discrete points.
%In addition they use different data structures to represent finite abstractions, and each one is designed with different theoretical guarantees and computational constraints in mind.

%This paper's role is to act as an intermediary. Flexibility is enabled through composability.
%These operations should be composable and powerful enough to recreate as many features of existing abstraction and synthesis toolboxes as possible.
%Doing so would make it easier for developers of control synthesis toolboxes to quickly write their own custom pipelines, heuristics, and optimizations depending on whether they value computational tractability or precision. Simultaneously, if a specific tool fails to work on a target application, users will be able to modify the tool without breaking any guarantees.

\subsection{Methodology}

%The primary contribution of this paper is to break apart the control synthesis pipeline into smaller modular components. This enables one to encode domain specific optimizations by ``hijacking" and modifying the control synthesis pipeline's topology, enabling one to 
%directly align the control synthesis algorithm with structural properties exhibited by the dynamics and specification.
%
%We use an augmented theory relational interfaces\cite{tripakis2011theory} so that it may be applied from abstraction-based controller synthesis.

\begin{figure}[t!]
\centering
\includegraphics[width=.8\columnwidth]{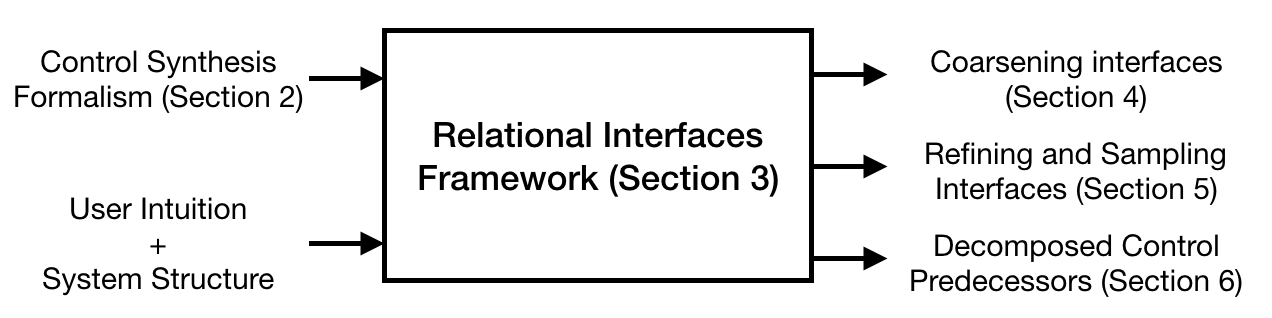}
\caption{\small 
By expressing many different techniques within a common framework, 
users are able to rapidly develop methods to exploit system structure in controller synthesis.
} \label{fig:outline}
\end{figure}

\Cref{fig:outline} depicts this paper's methodology and organization. The existing control synthesis formalism does not readily lend itself to algorithmic modifications that reflect and exploit structural properties in the system and specification.
We use the theory of relational interfaces \cite{tripakis2011theory} as a foundation and augment it to express control synthesis pipelines. Interfaces are used to represent both system models and constraints.
A small collection of atomic operators manipulates interfaces and is powerful enough to reconstruct many existing control synthesis pipelines. 
% Atomic operators allow one to construct computational pipelines in control synthesis by manipulating interfaces. While small, this collection is powerful enough to reconstruct many existing algorithms by stitching together atomic operators into composites.

One may also add new composite operators to encode desirable heuristics that exploit structural properties in the system and specifications.
The last three sections encode the techniques for abstraction-based control synthesis from \Cref{subsec:abcs} within the relational interfaces framework. By deliberately deconstructing those techniques, then reconstructing them within a compositional framework it was possible to identify implicit or unnecessary assumptions then generalize or remove them. It also makes the aforementioned techniques interoperable amongst themselves as well as future techniques.

Interfaces come equipped with a refinement partial order that formalizes when one interface abstracts another.
This paper focuses on preserving the refinement relation and sufficient conditions to refine discrete controllers back to concrete ones. Additional guarantees regarding completeness, termination, precision, or decomposability can be encoded, but impose additional requirements on the control synthesis algorithm and are beyond the scope of this paper.

\subsection{Contributions}

To our knowledge, the application of relational interfaces to robust abstraction-based control synthesis is new. 
The framework's building blocks consist of a collection of small, well understood operators that are nonetheless powerful enough to express many prior techniques.
Encoding these techniques as relational interface operations forced us to simplify, formalize, or remove implicit assumptions in existing tools. The framework also exhibits numerous desirable features.
\begin{enumerate}%
\item It enables compositional tools for control synthesis by leveraging a theoretical foundation with compositionality built into it. This paper showcases a principled methodology to seamlessly combine the methods in \Cref{subsec:abcs}, as well as construct new techniques.
\item It enables a declarative approach to control synthesis by enforcing a strict separation between the high level algorithm from its low level implementation. We rely on the availability of an underlying data structure to encode and manipulate predicates. Low level predicate operations, while powerful, make it easy to inadvertently violate the refinement property. Conforming to the relational interface operations minimizes this danger.
\end{enumerate}
This paper's first half is domain agnostic and applicable to general robust control synthesis problems. The second half applies those insights to the finite abstraction approach to control synthesis.
A smaller Dubins vehicle example is used to showcase and evaluate different techniques and their computational gains, compared to the unoptimized problem.
\ifcav In an extended version of this paper available at \cite{redax}, a 6D lunar lander example leverages all techniques in this paper and introduces a few new ones.
\else
A 6D lunar lander example is included in the appendix which leverages all of the techniques in this paper collectively.
\fi

%\todo{Contrast with SLUGS?: The GR(1) synthesis tool SLUGS is oriented towards discrete systems and fixed specifications. Plugins to bias results to incorporate favorable properties. More about different solutions to different theoretical problems rather than about computational benefits.}

%Sections \ref{sec:interf}-\ref{sec:core} introduce relational interfaces as models for sets and dynamics as well as atomic operators to transform interfaces. \Cref{sec:refinementpo} introduces the refinement partial order, and Sections \ref{sec:quantinter}-\ref{sec:abs} show how to traverse it vertically by coarsening and refinement. Finite abstraction is a combination of both. The lunar lander example in \Cref{sec:examples} showcases different techniques to cope with the system's high dimensionality.

\subsection{Notation}

%Let $\true$ denote logical true and $\false$ denote false. Operators $\NOT, \AND, \OR$ respectively represent negation, conjunction, and disjunction. The implication $ a \implies b $ is shorthand for the formula $\NOT a \OR b$.

Let $=$ be an \textit{assertion} that two objects are mathematically equivalent; as a special case `$\sequiv$' is used when those two objects are sets. In contrast, the operator `$==$'  \textit{checks} whether two objects are equivalent, returning true if they are and false otherwise. A special instance of `$==$' is logical equivalence `$\lequiv$'. 

Variables are denoted by lower case letters. Each variable $v$ is associated with a domain of values $\dom{v}$ that is analogous to the variable's type. A composite variable is a set of variables and is analogous to a bundle of wrapped wires. 
From a collection of variables $\V_1, \ldots, \V_M$ a composite variable $\V$ can be constructed by taking the union $\V \sequiv \V_1 \cup \ldots \cup \V_M$ and the domain $\dom{\V} \equiv \prod_{i=1}^M \dom{\V_i}$.
Note that the variables $\V_1, \ldots, \V_M$ above may themselves be composite. 
As an example if $v$ is associated with a $M$-dimensional Euclidean space $\mathbb{R}^M$, then it is a composite variable that can be broken apart into a collection of atomic variables $v_1, \ldots, v_M$ where $\dom{v_i} \sequiv \mathbb{R}$ for all $i \in \{1,\ldots, M\}$.
The technical results herein do not distinguish between composite and atomic variables.  %\todo{Variables are not unbounded to a specific assignment.}

Predicates are functions that map variable assignments to a Boolean value. 
%Boolean valued expressions like ``$x \in \{4,5,12\}$" and ``$y == \sin(x)$" are predicates. The variables contained in those expressions are unassigned in the sense that they are not associated with a single value. 
Predicates that stand in for expressions/formulas are denoted with capital letters.
%and are often written with the variables that appear within them, e.g. a predicate $P$ over variables $v, w$ can stand in for the expression $``v \leq w"$.
%Predicates can construct sets via set builder notation. A single predicate can instantiate different sets if the domains differ, e.g. $\{ x \in \dom{\XX} | P(x)\}$ and $\{(x,y) \in \dom{\XX} \times \dom{\YY} | P(x)\}$ are distinct sets but are associated with the same predicate.
%Boolean operations can be applied to a predicate's Boolean output to construct new predicates. The negated predicate $\neg P(v)$ is true for an assignment to $v$ if and only if $P(v)$ is false. 
%The domain of a predicate obtained via a binary operation is the union of the two variable domains, e.g., conjunction $P(v) \AND Q(w)$ yields a predicate $(P \AND Q)(v,w)$.
Predicates $P$ and $Q$ are logically equivalent (denoted by $P \lequiv Q$) if and only if $P \implies Q$ and $Q \implies P$ are true for all variable assignments.
The universal and existential quantifiers $\forall$  and $\exists$ eliminate variables and yield new predicates.
Predicates  $\equant{\W}P$ and $\uquant{\W}P$ do not depend on $\W$.
%An assignment $v \in \dom{v}$ satisfies $\equant{\W}P$ if and only if there exists an assignment $w \in \dom{w}$ such that $P$ evaluates to true. 
%Similarly, an assignment to $v$ satisfies $\uquant{\W}P$ if and only if $P(v,w)$ evaluates to true for all assignments $w \in \dom{w} $.
%Applying DeMorgan's law yields the identities $\neg \equant{\W}P \lequiv \uquant{\W}(\neg P)$ and $\neg \uquant{\W}P \lequiv \equant{\W}(\neg P)$.
%If the variable to be eliminated does not exist in the predicate, then the same predicate is returned. 
If $\W$ is a composite variable $\W \sequiv \W_1 \cup \ldots \cup \W_N$ then $\equant{\W}P$ is simply a shorthand for $\equant{\W_1}\ldots\equant{\W_N} P$.

%\todo{Predicate representation as BDDs}

%\begin{remark}
%Predicate variables are omitted when necessary in order to avoid bloated expressions and notation overhead. For example, $P(v,w)$ can simply be denoted by $P$ when clear from context that it is associated with $v$ and $w$.
%Moreover, if some variables are composite then the notation can be expanded. That is, if $v \vequiv v_1 \cup \ldots \cup v_M$ then $P(v_1, \ldots, v_M, w)$ and $P(v,w)$ represent the same predicate.
%\end{remark}

%\todo{In finite domains, many implementations available for storing predicates and applying operations. They can be viewed as tables in a database, a tree, boolean valued circuit, bdd, etc.}

\section{Control Synthesis for a Motivating Example}

As a simple, instructive example consider a planar Dubins vehicle that is tasked with reaching a desired location. Let $x = \{p_x, p_y, \theta\}$ be the collection of state variables, $u = \{v, \omega\}$ be a collection input variables to be controlled, $x^+ = \{p_x^+, p_y^+, \theta^+\}$ represent state variables at a subsequent time step, and $L = 1.4$ be a constant representing the vehicle length.
The constraints
\begin{align}
p_x^+ &== p_x + v \cos(\theta) \label{eqn:dubx} \tag{$F_x$}\\
p_y^+ &== p_y + v \sin(\theta) \tag{$F_y$} \\
\theta^+ &== \theta + \dfrac{v}{L} \sin(\omega) \label{eqn:dubtheta} \tag{$F_\theta$}
\end{align} 
characterize the discrete time dynamics.
The continuous state domain is $\dom{x} \sequiv [-2,2] \times [-2,2] \times [-\pi,\pi)$, where the last component is periodic so $-\pi$ and $\pi$ are identical values. The input domains are $\dom{v} \sequiv \{0.25, 0.5\}$ and $\dom{\omega} \sequiv \{-1.5, 0, 1.5\}$
%The position constraint $p \in [-8,8]$ encodes the collision free region.
%Let this region be represented via with the predicate $T$.
%We constrain the drone's movement to be along a single vertical direction and is modeled by a discrete-time double integrator; it is easier to visualize low dimensional state spaces and dynamics.

%The transition relation is encoded by two predicates where $p$ represents position, $v$ represents velocity, and parameters satisfy $T = 0.2$, $k = .1$, $g=9.8$.
%\begin{align}
%p^+ &== p + Tv  \\
%v^+ &== v + Ta - kT\text{sgn}(v)v^2 - Tg
%\end{align}
%Let $x = \{p, v\}$ and $u = \{a\}$.

Let predicate $F = F_x \AND F_y \AND F_\theta$ represent the monolithic system dynamics. Predicate $T$ depends only on $x$ and represents the target set $[-0.4,0.4] \times [-0.4,0.4] \times [-\pi,\pi)$, encoding that the vehicle's position must reach a square with any orientation.
%Let discrete time  \footnote{Control synthesis toolboxes translate continuous time systems to discrete time by fixing a sampling time step and imposing controllable inputs to have a zero order hold \cite{gunther2}.} system dynamics be encoded as an input-output interface $F(x \cup u, x^+)$. Variable $x$ represents a current state, $u$ represents a controllable input, and $x^+$ represents the state at a subsequent time step.
%Let $Z(x^+, \emptyset)$ be a sink interface representing a target set of next states.
Let $Z$ be a predicate that depends on variable $x^+$ that encodes a collection of states at a future time step.
 \Cref{eqn:origcpre} characterizes the robust controlled predecessor, which takes $Z$ and computes the set of states from which there exists a non-blocking assignment to $u$ that guarantees $x^+$ will satisfy $Z$, despite any non-determinism contained in $F$. The term $\equant{x^+}F$ prevents state-control pairs from blocking, while $\uquant{x^+}(F \implies Z)$ encodes the state-control pairs that guarantee satisfaction of $Z$.
\begin{align}
\CPRE(F, Z) = \equant{u} ( \equant{x^+} F \AND \uquant{x^+}(F \implies Z)). \label{eqn:origcpre}
\end{align}
%Note that $F$ and $Z$ share variable(s) $x^+$.
%This can be visualized as a connection between blocks at the top of \Cref{fig:cpre}. The rest of the figure depicts how $\CPRE$ could be viewed as applying a sequence of smaller interface transformations, which we now introduce.

%Let $T$ denote a predicate that encodes a collision-free region for the drone.
The controlled predecessor is used to solve safety and reach games. We can solve for a region for which the target $T$ (respectively, safe set $S$) can be reached (made invariant) via an iteration of an appropriate $\code{reach}$ ($\code{safe}$) operator. Both iterations are given by:
\begin{align}
\text{Reach Iter:} \qquad Z_0 &= \false %\label{reachgame:init} \\ 
\qquad Z_{i+1} = \code{reach}(F,Z_i,T) = \CPRE(F, Z_{i}) \OR T. \label{reachgame:iter}\\% \tag{Reach}
\text{Safety Iter:}\qquad Z_0 &= S %\label{safegame:init} \\ 
\qquad Z_{i+1} = \code{safe}(F, Z_i, S)= \CPRE(F, Z_{i}) \AND S. \label{safegame:iter} %\tag{Safe}
\end{align}
The above iterations are not guaranteed to reach a fixed point in a finite number of iterations, except under certain technical conditions \cite{Tabu}.
\begin{wrapfigure}{r}{2.1in}
\centering
\includegraphics[width=2in]{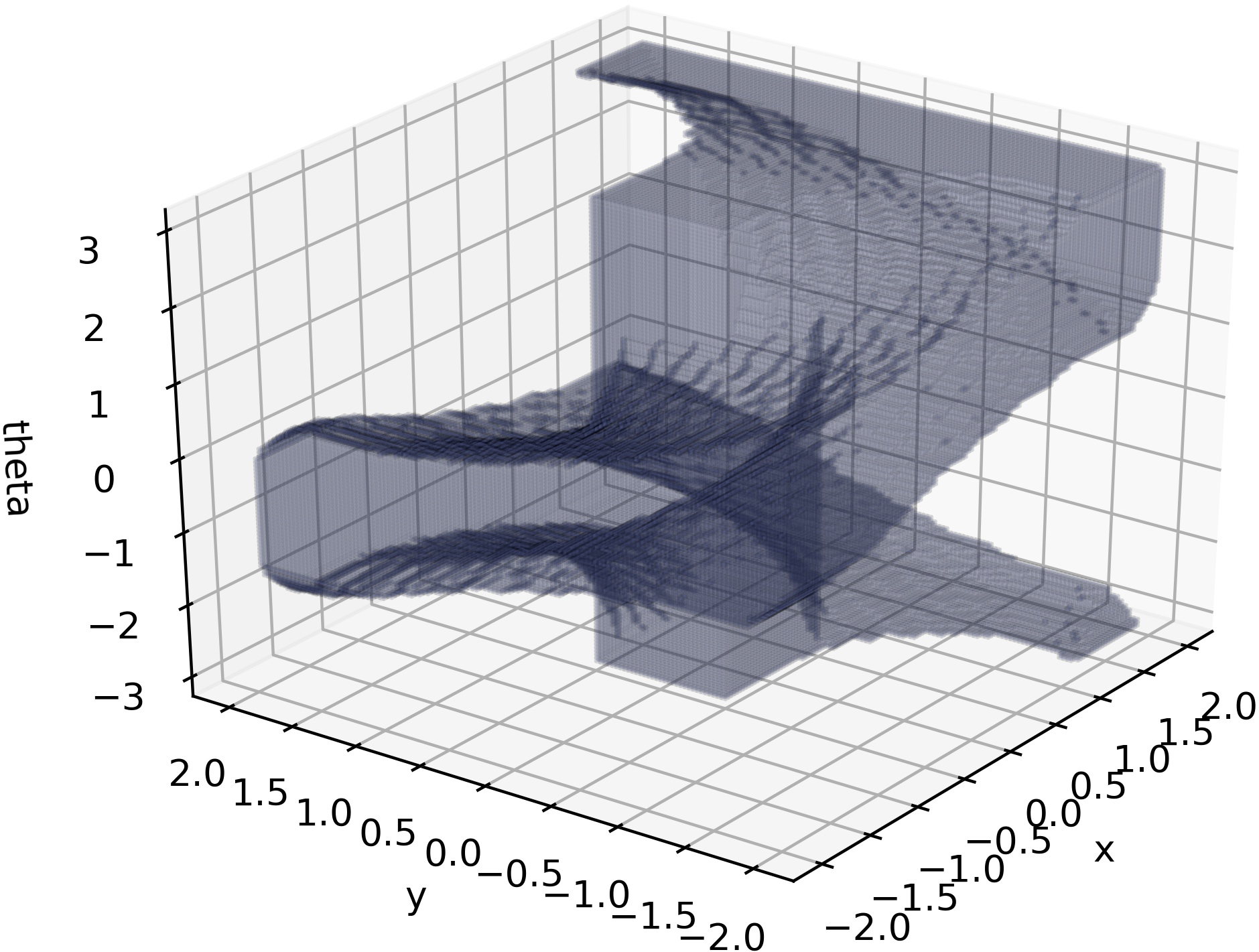}
\caption{\small Approximate solution to the Dubins vehicle reach game visualized as a subset of the state space.} \label{dubinsapprox}
\end{wrapfigure}
\Cref{dubinsapprox} depicts an approximate region where the controller can force the Dubins vehicle to enter $T$.
We showcase different improvements relative to a base line script used to generate \Cref{dubinsapprox}.
A toolbox that adopts this paper's framework is being actively developed and is open sourced at \cite{redax}. It is written in \texttt{python 3.6} and uses the \texttt{dd} package as an interface to \texttt{CUDD}\cite{Somenzi2015}, a library in \texttt{C/C++} for constructing and manipulating binary decision diagrams (BDD).
All experiments were run on a single core of a 2013 Macbook Pro with 2.4GHz Intel Core i7 and 8GB of RAM.

The following section uses relational interfaces to represent the controlled predecessor $\CPRE(\cdot)$ and iterations \pref{reachgame:iter} and  \pref{safegame:iter} as a computational pipeline. Subsequent sections show how modifying this pipeline leads to favorable theoretical properties and computational gains.
%\begin{minted}{python}
%class SafetyGame(F, W):
%
%	def __init__(F, W, var):
%		a
%		
%	
%	def CPRE(Z):
%		# Computes 	
%		return (F.exist(x+) & (!F | Z).forall(x+)).exist(u)
%		
%	def solve(Z):
%		ZZ = CPRE(Z)
%		if ZZ = Z:
%			return ZZ
%		else:
%			return CPRE(Z)
%		
%		
%\end{minted}

\section{Relational Interfaces} \label{sec:interf}

Relational interfaces are predicates augmented with annotations about each variable's role as an input or output \footnote{Relational interfaces closely resemble assume-guarantee contracts \cite{NuzzoThesis}; we opt to use relational interfaces because inputs and outputs play a more prominent role.}. They abstract away a component's internal implementation and only encode an input-output relation. 
\begin{definition}[Relational Interface \cite{tripakis2011theory}]
An interface $M(\II,\OO)$ consists of a predicate $M$ over a set of input variables $\II$ and output variables $\OO$.
\end{definition}
For an interface $M(i,o)$, we call $(i,o)$ its input-output \textit{signature}. 
An interface is a sink if it contains no outputs and has signature like $(i, \emptyset)$, and a source if it contains no inputs like $(\emptyset, o)$. Sinks and source interfaces can be interpreted as sets whereas input-output interfaces are relations.
Interfaces encode relations through their predicates and can capture features such as non-deterministic outputs or blocking (i.e., disallowed, error) inputs. 
A system blocks for an input assignment if there does not exist a corresponding output assignment that satisfies the interface relation.
Blocking is a critical property used to declare \textit{requirements};  sink interfaces impose constraints by modeling constrain violations as blocking inputs.
Outputs on the other hand exhibit non-determinism, which is treated as an \textit{adversary}. When one interface's outputs are connected to another's inputs, the outputs seek to cause blocking whenever possible.
%For interfaces encoding a transition relation, errors could signify that the state has exited the region where the model is valid. 

%While the general theory of relational interfaces contained in \cite{tripakis2011theory} can accommodate stateful interfaces, all interfaces in this paper will be stateless.
%The theory and motivation behind relational interfaces closely resembles those of assume-guarantee contracts \cite{NuzzoThesis}; we opt for relational interfaces because the roles of inputs and outputs play a more prominent role.

%State transitions in control synthesis will.
%A similar modeling assumption is made in \cite{kim2018abstractions}. The way the different roles are distinguished is through how the interfaces are manipulated and transformed through operators.
%Interface inputs and outputs have no intrinsic meaning within the context of controller synthesis. Inputs can be used to represent disturbances, controllable inputs, and current states while next states will be represented by interface outputs.

\subsection{Atomic and Composite Operators} \label{sec:core}

%\begin{table}
%\centering
%\begin{tabular}{|c|c|c|}
%\hline 
%Atomic Operators  & Predicate  & Interface \\
%  & Representation & Signature \\ \hline
%$\ohide(w,F(i,o))$ & $\equant{w}F$ & $(i, o \setminus w)$ \\ \hline
%$\ihide(w,F(i,\emptyset))$ & $\equant{w}F$  & $(i \setminus w, \emptyset)$ \\ \hline
%$\comp(F_1(i_1,o_1), F_2(i_2, o_2))$ & $F_1 \AND F_2 \AND \uquant{io_{12}} (F_1 \implies \NB{F_2})$ & $((i_1 \cup i_2) \setminus io_{12}, o_1 \cup o_2)$  \\
%& where $io_{12} = i_2 \cap o_1$ &  \\ \hline
%$\refine(F_1(i,o),F_2(i,o))$ & $(\NB{F_1} \OR \NB{F_2}) $ & $(i,o)$\\
%& $\AND (\NB{F_1} \implies F_1) \AND (\NB{F_2} \implies F_2) $ &  \\ \hline 
%\end{tabular}
%\end{table}

Operators are used to manipulate interfaces by taking interfaces and variables as inputs and yielding another interface.
%Operators can also be composed to construct new ones.
We will show how the controlled predecessor $\CPRE(\cdot)$ in \pref{eqn:origcpre} can be constructed by composing operators appearing in \cite{tripakis2011theory} and one additional one.
%\todo{Control synthesis toolboxes temporally discretize continuous time systems by fixing a sampling time step and imposing controllable inputs to have a zero order hold \cite{gunther2}.}
The first, output hiding, removes interface outputs.%, such as those that were shared internal variables before composition relabeled them as outputs.
\begin{definition}[Output Hiding \cite{tripakis2011theory}] \label{def:varhiding}
Output hiding operator $\ohide(w, F)$ over interface $F(\II, \OO)$ and outputs $w$ yields an interface with signature $(\II,\OO \setminus w)$.
\begin{align}
\ohide(w, F) = \equant{w}F
\end{align}
\end{definition}
Existentially quantifying out $w$ ensures that the input-output behavior over the unhidden variables is still consistent with potential assignments to $w$.
The operator $\NB{\cdot}$ is a special variant of $\ohide(\cdot)$ that hides all outputs, yielding a sink encoding all non-blocking inputs to the original interface.
%In the context of \Cref{fig:cpre}, $\ohide(x^+, \comp(F,Z))$ is a sink interface $(x \cup u, \emptyset)$ signifying the set of all state-input pairs for which $Z$ will be satisfied under dynamics $F$. 

\begin{definition}[Nonblocking Inputs Sink] \label{def:NB}
Given an interface $F(\II,\OO)$, the nonblocking operation \NB{F} yields a sink interface with signature $(i, \emptyset)$ and predicate $\NB{F} = \equant{\OO}F$.
%\begin{align}
%\NB{F} = \equant{\OO}F.
%\end{align}
If $F(i, \emptyset)$ is a sink interface, then $\NB{F} = F$ yields itself. If $F(\emptyset, o)$ is a source interface, then $\NB{F} = \false$ if and only if $F \lequiv \false$; otherwise $\NB{F} = \true$. 
\end{definition}

The interface composition operator takes multiple interfaces and ``collapses" them into a single input-output interface. It can be viewed as a generalization of function composition in the special case where each interface encodes a total function (i.e., deterministic output and inputs never block).
%\begin{figure}
%\centering
%\includegraphics[width=.5\columnwidth]{figs/cpresequence.png}
%\caption{\small Control predecessor as a sequence of interface operations. Interface $F(x \cup u, x^+)$ represents system dynamics and sink interface $Z(x^+, \emptyset)$ represents a target set of states. \todo{Pseudocode in python}} \label{fig:cpre}
%\end{figure}

\newcommand\IO[1]{io_{#1}}
\begin{definition}[Interface Composition \cite{tripakis2011theory}] 
Let $F_1(\II_1, \OO_1)$ and $F_2(\II_2, \OO_2)$ be interfaces with disjoint output variables $\OO_1 \cap \OO_2 \sequiv \emptyset$ and $\II_1 \cap \OO_2 \sequiv \emptyset$ which signifies that $F_2$'s outputs may not be fed back into $F_1$'s inputs.
Define new composite variables
\begin{align}
\IO{12} &\sequiv \OO_1 \cap \II_2\\
\II_{12} &\sequiv (\II_1 \cup \II_2) \setminus \IO{12} \\
\OO_{12} &\sequiv \OO_1 \cup \OO_2. %\cup \IO{12}
\end{align}
Composition $\comp(F_1, F_2)$ is an interface with signature $(i_{12}, o_{12})$ and predicate
\begin{align}
F_1 \AND F_2  \AND \uquant{\OO_{12}}(F_1 \implies \NB{F_2}). \label{eq:compositionpredicate}
\end{align}
Interface subscripts may be swapped if instead $F_2$'s outputs are fed into $F_1$.
\label{def:modcomposition}
\end{definition}

Interfaces $F_1$ and $F_2$ are composed in parallel if $io_{21} \sequiv \emptyset$ holds in addition to $io_{12} \sequiv \emptyset$. \Cref{eq:compositionpredicate} under parallel composition reduces to $F_1 \AND F_2$ (Lemma 6.4 in \cite{tripakis2011theory}) and $\comp(\cdot)$ is commutative and associative. If $io_{12} \not\equiv \emptyset$, then they are composed in series and the composition operator is only associative. 
%\Cref{fig:cpre} depicts the series composition of $F$ and $Z$.
Any acyclic interconnection can be composed into a single interface by systematically applying \Cref{def:modcomposition}'s binary composition operator.
Non-deterministic outputs are interpreted to be \textit{adversarial}.
Series composition of interfaces has a built-in notion of robustness to account for $F_1$'s non-deterministic outputs and blocking inputs to $F_2$ over the shared variables $io_{12}$.
%The last term in \Cref{eq:compositionpredicate} is a predicate over the expanded input set $i_{12}$ and deals with blocking behaviors under series composition.
%The series composition changes the role of variables $io_{12} \subseteq i_2$ from inputs to outputs.
The term $\uquant{\OO_{12}}(F_1 \implies \NB{F_2})$ in \Cref{eq:compositionpredicate} is a predicate over the composition's input set $i_{12}$. It ensures that if a potential output of $F_1$ may cause $F_2$ to block, then $\comp(F_1, F_2)$ must preemptively block.
%\begin{align}
%\NB{\comp(F_1, F_2)} = \equant{o_{12}} (F_1 \AND F_2) \AND \uquant{o_{12}} (F_1 \implies \NB{F_2})
%\end{align}

%This is one of many instances of composite operators being constructed from smaller primitives.
%Applying a nonblocking operator on every single output yields a sink interface that encodes the interface's nonblocking inputs.
The final atomic operator is input hiding, which may only be applied to sinks. If the sink is viewed as a constraint, an input variable is ``hidden" by an angelic environment that chooses an input assignment to satisfy the constraint. This operator is analogous to projecting a set into a lower dimensional space.
\begin{definition} [Hiding Sink Inputs]
Input hiding operator $\ihide(w,F)$ over sink interface $F(i, \emptyset)$ and inputs $w$ yields an interface with signature $(i \setminus w, \void)$.
\begin{align}
\ihide(w, F) = \equant{w} F
\end{align}
%yields an interface with signature . 
\end{definition}
Unlike the composition and output hiding operators, this operator is not included in the standard theory of relational interfaces \cite{tripakis2011theory} and was added to encode a controller predecessor introduced subsequently in \Cref{eqn:compositecpre}.

%\begin{table}
%\centering
%\begin{tabular}{|c|c|c|}
%\hline
%Composite  & Atomic  & Interface \\ 
%Operators& Representation & Signature \\ \hline 
%$\NB{F(i,o)}$ & $\equant{o}F$  & $(i,\emptyset)$ \\ \hline
%$\CPRE(F(x \cup u, x^+),Z(x^+, \emptyset))$ & $\ihide(u, \ohide(x^+, \comp(F,Z)))$ & $(x, \emptyset)$ \\ \hline
%$\incoarsen(F(i,o), Q(\abs i, i))$ & $\ohide(i, \comp(Q(\abs i, i), F) )$ & $(\abs i,o)$ \\ \hline
%$\outcoarsen(F(i,o), Q(o, \abs o))$ & $ \ohide(o, \comp( F, Q(o, \abs o)))$ & $(i, \abs o)$ \\ \hline
%\end{tabular}
%\end{table}

\subsection{Constructing Control Synthesis Pipelines}

The robust controlled predecessor (\ref{eqn:origcpre}) can be expressed through operator composition.
%Note that operator composition and interface composition are distinct within this framework.
\begin{proposition} \label{prop:cpreop}
The controlled predecessor operator \pref{eqn:compositecpre} % $\CPRE(F,Z)$ 
%\pref{eqn:compositecpre}.
yields a sink interface with signature $(x, \emptyset)$ and predicate equivalent to the predicate in \pref{eqn:origcpre}.
\begin{align}
\CPRE(F, Z) &= \ihide(u, \ohide(x^+, \comp(F,Z))). \label{eqn:compositecpre}
\end{align}
\end{proposition}
\ifcav
The simple proof is provided in the extended version at \cite{redax}.
\else
\begin{proof}
Applying the definitions of $\comp$, $\ihide$, and $\ohide$ yields the expression 
\begin{align}
\equant{u}\equant{x^+}(F \AND Z \AND \uquant{x^+}(F \implies Z)))
\end{align}
One can safely move the $\equant{x^+}$ inside the parenthesis because $\uquant{x^+}(F \implies Z)$ is a predicate that is independent of $x^+$, yielding
\begin{align}
\equant{u}(\equant{x^+}(F \AND Z) \AND \uquant{x^+}(F \implies Z)))
\end{align}
To show equivalence of the above expressions with \Cref{eqn:origcpre}, we simply need to show equivalence of the following two predicates that depend on $x$ and $u$:
\begin{align}
\equant{x^+}(F \AND Z) \AND \uquant{x^+}(F \implies Z) \label{eqn:xu1}\\
\equant{x^+} F \AND \uquant{x^+}(F \implies Z). \label{eqn:xu2}
\end{align}
It is easy to see that \pref{eqn:xu1} implies \pref{eqn:xu2} because $\equant{x^+}(F\AND Z)$ implies $\equant{x^+}F$. To show the reverse, suppose \pref{eqn:xu2} is satisfied for a pair $x$ and $u$. Any $x^+$ chosen to satisfy $\equant{x^+}F$ must also satisfy the constraint imposed by the sink $Z$. Otherwise, the clause $\uquant{x^+}(F \implies Z)$ would be violated, contradicting satisfaction of \pref{eqn:xu2}. Therefore, \pref{eqn:xu1} and \pref{eqn:xu2} are equivalent which completes the proof.
\end{proof}
\fi
 %provided in \Cref{cprepropproof} of the appendix.
\Cref{prop:cpreop} signifies that controlled predecessors can be interpreted as an instance of robust composition of interfaces, followed by variable hiding.
It can be shown that $\code{safe}(F,Z, S) = \comp(\CPRE(F, Z), S)$ because $S(x, \emptyset)$ and $\CPRE(F, Z)$ would be composed in parallel. \footnote{Disjunctions over sinks are required to encode $\code{reach}(\cdot)$. This will be enabled by the shared refinement operator defined in \Cref{def:refinement}.}
%\begin{wrapfigure}{r}{2.5in}
\begin{figure}[!t]
\centering
\includegraphics[width=.5\columnwidth]{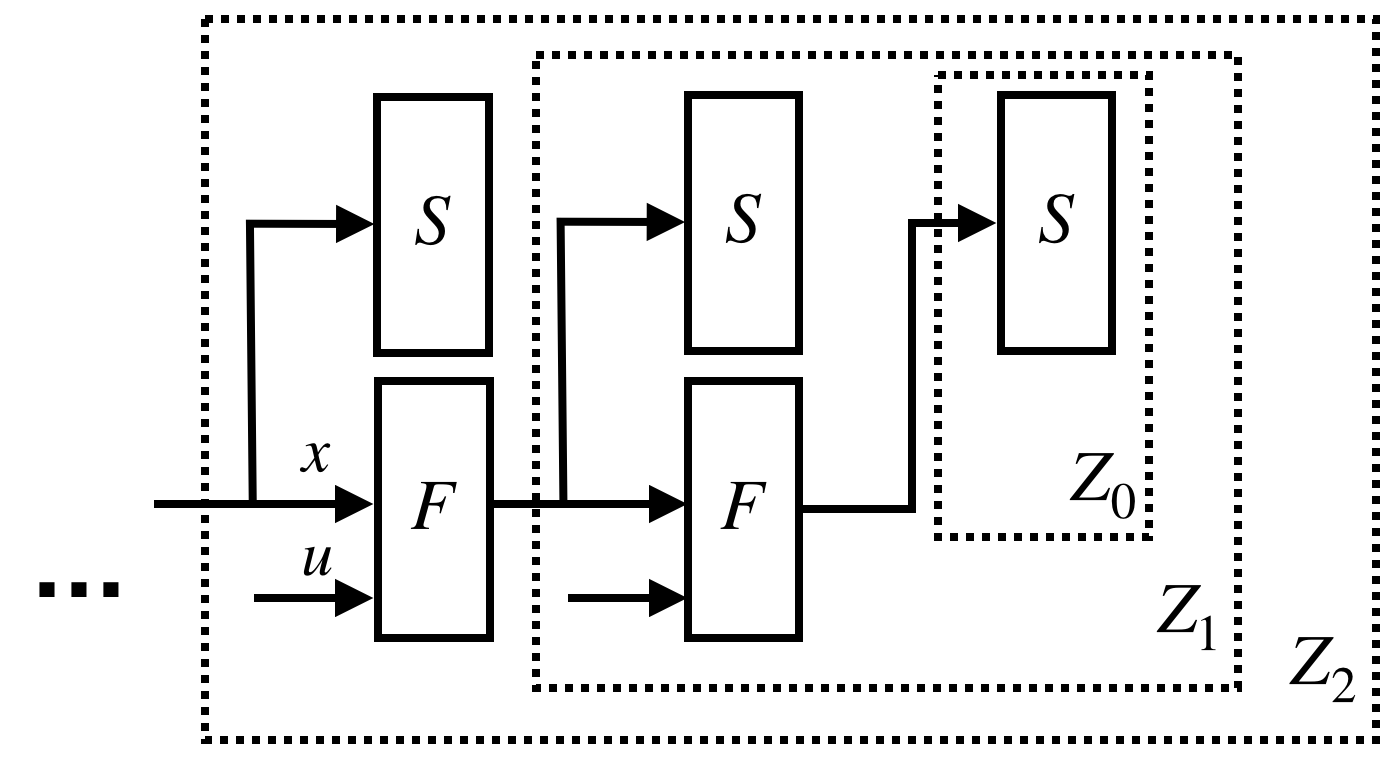}
\caption{\small Safety control synthesis iteration \pref{safegame:iter} depicted as a sequence of sink interfaces.} \label{fig:cprepipeline}
%\end{wrapfigure}
\end{figure}
\Cref{fig:cprepipeline} shows a visualization of the safety game's fixed point iteration from the point of view of relational interfaces.
Starting from the right-most sink interface $S$ (equivalent to $Z_0$) the iteration \pref{safegame:iter} constructs a sequence of sink interfaces $Z_1, Z_2,...$ encoding relevant subsets of the state space.
The numerous $S(x, \emptyset)$ interfaces impose constraints and can be interpreted as monitors that raise errors if the safety constraint is violated.
%The controlled predecessor $\CPRE(F,Z)$ collapses $Z_0$ and $F$ down into another sink interface $Z_1$. This sink, when composed with $F$, yields another sink interface over the variables $x \cup u$.

\subsection{Modifying the Control Synthesis Pipeline}

\Cref{eqn:compositecpre}'s definition of $\CPRE(\cdot)$ is oblivious to the domains of variables $x,u$, and $x^+$. This generality is useful for describing a problem and serving as a blank template.
%The most common reason to modify the pipeline is to computational tractability.
Whenever problem structure exists, pipeline modifications refine the general algorithm into a form that reflects the specific problem instance.
They also allow a user to inject implicit preferences into a problem and reduce computational bottlenecks or to refine a solution.
%They also can explicitly capture structure, whenever it exists, and mold an algorithm to reflect it.
%alternatively to inject it.
%Only when a specific structural property in the dynamics or the specification.
%\todo{Make more concrete. Domain specific issues like termination. Structure is often domain specific. Modifications are domain specific.}
%The tools in  implement the abstraction and synthesis steps; some are augmented with techniques to mitigate the computational burdens of synthesizing controllers for systems with high dimensional continuous state spaces.
The subsequent sections apply this philosophy to the abstraction-based control techniques from \Cref{subsec:abcs}:
\begin{itemize}
\item \Cref{sec:refinementpo}: Coarsening interfaces reduces the computational complexity of a problem by throwing away fine grain information. The synthesis result is conservative but the degree of conservatism can be modified.
\item \Cref{sec:abs}: Refining interfaces decreases result conservatism. Refinement in combination with coarsening allows one to dynamically modulate the complexity of the problem as a function of multiple criteria such as the result granularity or minimizing computational resources. 
\item \Cref{sec:decomppre}: If the dynamics or specifications are decomposable then the control predecessor operator can be broken apart to refect that decomposition.
\end{itemize}
These sections do more than simply reconstruct existing techniques in the language of relational interfaces. 
They uncover some implicit assumptions in existing tools and either remove them or make them explicit.
Minimizing the number of assumptions ensures applicability to a diverse collection of systems and specifications and compatibility with future algorithmic modifications.

\section{Interface Abstraction via Quantization} \label{sec:refinementpo}

A key motivator behind abstraction-based control synthesis is that computing the game iterations from \Cref{reachgame:iter} and \Cref{safegame:iter} exactly is often intractable for high-dimensional nonlinear dynamics. Termination is also not guaranteed.
%By making $F$ and $Z$ finite, the iterations are guaranteed to terminate.
Quantizing (or ``abstracting") continuous interfaces into a finite counterpart ensures that each predicate operation of the game terminates in finite time but at the cost of the solution's precision. Finer quantization incurs a smaller loss of precision but can cause the memory and computational requirements to store and manipulate the symbolic representation to exceed machine resources. 

This section first introduces the notion of interface abstraction as a refinement relation. We define the notion of a quantizer and show how it is a simple generalization of many existing quantizers in the abstraction-based control literature. Finally, we show how one can inject these quantizers anywhere in the control synthesis pipeline to reduce computational bottlenecks.
%To counter this, one may inject coarser quantizers directly into the synthesis pipeline to simplify computations and reduce memory usage whenever it becomes apparent that the fixed point computation's will require more computational resources than are available.

%\todo{Why quantize at all? Throws away information and reduces the complexity of the problem. When you add to a problem, you are declaring that the problem is too hard to solve and you are willing to take approximations.}

%One such instance is an quantizer that maps from $[0,1] \subset \mathbb{R}$ to sets in a finite cover.
%\begin{definition}[Quantizer Interface]
%\todo{quantizer blocks}
%A quantizer interface in practice is an interface that maps elements from a higher cardinality input set into a lower cardinality output set. 

%\end{definition}

%\Cref{subsec:abstheory}

\subsection{Theory of Abstract Interfaces} \label{subsec:abstheory}
While a controller synthesis algorithm can analyze a simpler model of the dynamics, the results have no meaning unless they can be extrapolated back to the original system dynamics. The following interface refinement condition formalizes a condition when this extrapolation can occur.

\begin{definition}[Interface Refinement \cite{tripakis2011theory}] \label{def:relabs}
Let $F(i,o)$ and $\abs F(\abs i, \abs o)$ be interfaces. $\abs F$ is an abstraction of $F$ if and only if $i \sequiv \abs i$, $o \sequiv \abs o$, and
%\begin{align}
%(\quant{\II}(i,\abs i) \AND \NB{\abs M}(\abs i)) \implies \NB{M}(i) \label{eqn:NB}
%\end{align}
\begin{align}
\NB{\abs F} &\implies \NB{F} \label{eqn:NB}\\
\left( \NB{\abs F} \AND F\right) &\implies
  \abs F
 \label{eqn:superset}
\end{align} 
are valid formulas. This relationship is denoted by $\abs F \modabs F$.
\end{definition}

\Cref{def:relabs} imposes two main requirements between a concrete and abstract interface. \Cref{eqn:NB} encodes the condition where if $\abs F$ accepts an input, then $F$ must also accept it; that is, the abstract component is more aggressive with rejecting invalid inputs. Second, if both systems accept the input then the abstract output set is a superset of the concrete function's output set.
The abstract interface is a conservative representation of the concrete interface because the abstraction accepts fewer inputs and exhibits more non-deterministic outputs.
If both the interfaces are sink interfaces, then $\abs F \modabs F$ reduces down to $\abs F \subseteq F$ when $F, \abs F$ are interpreted as sets. If both are source interfaces then the set containment direction is flipped and $\abs F \modabs F$ reduces down to $F \subseteq \abs F$.

\begin{figure}
\centering
\includegraphics[width=.9\columnwidth, height = 1.35in]{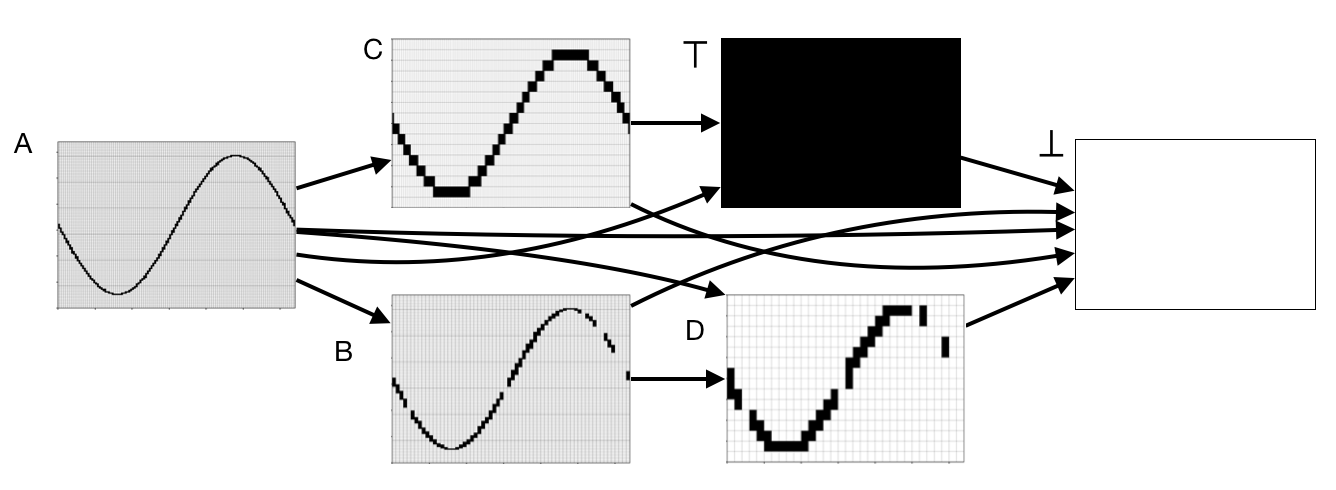}
\caption{\small Example depiction of the refinement partial order. Each small plot on the depicts input-output pairs that satisfy an interface's predicate. Inputs (outputs) vary along the horizontal (vertical) axis. Because $B$ blocks on some inputs but $A$ accepts all inputs $B \modabs A$. Interface $C$ exhibits more output non-determinism than $A$ so $C \modabs A$. Similarly $D \modabs B$, $D \modabs C$ , $\true \modabs C$, etc. Note that $B$ and $C$ are incomparable because $C$ exhibits more output non-determinism and $B$ blocks for more inputs. The false interface $\false$ is a universal abstraction, while $\true$ is incomparable with $B$ and $D$.
%\Cref{tbl:coarsen} later shows how certain interfaces were obtained through coarsening.
} \label{fig:order}
\end{figure}

The refinement relation satisfies the required reflexivity, transitivity, and antisymmetry properties to be a partial order \cite{tripakis2011theory} and is depicted in \Cref{fig:order}.
This order has a bottom element $\false$ which is a universal abstraction. Conveniently, the bottom element $\false$ signifies both boolean false and the bottom of the partial order. This interface blocks for every potential input.
In contrast, Boolean $\true$ plays no special role in the partial order. While $\true$ exhibits totally non-deterministic outputs, it also accepts inputs. A blocking input is considered ``worse" than non-deterministic outputs in the refinement order.
The refinement relation $\modabs$ encodes a direction of conservatism such that any reasoning done over the abstract models is sound and can be generalized to the concrete model.
\begin{theorem}[Informal Substitutability Result \cite{tripakis2011theory}]
For any input that is allowed for the abstract model, the output behaviors exhibited by an abstract model contains the output behaviors exhibited by the concrete model.
\end{theorem}
If a property on outputs has been established for an abstract interface, then it still holds if the abstract interface is replaced with the concrete one. Informally, the abstract interface is more conservative so if a property holds with the abstraction then it must also hold for the true system.
All aforementioned interface operators preserve the properties of the refinement relation of \Cref{def:relabs}, in the sense that they are monotone with respect to the refinement partial order. 

\begin{theorem}[Composition Preserves Refinement \cite{tripakis2011theory}]
Let $\abs A \modabs A$ and $\abs B \modabs B $. If the composition is well defined, then $\code{comp}(\abs A, \abs B) \modabs \code{comp}(A,B)$. \label{thm:comppresref}
\end{theorem}

\begin{theorem}[Output Hiding Preserves Refinement \cite{tripakis2011theory}]
If $A \modabs B$, then $\ohide(w,A) \modabs \ohide(w,B)$ for any variable $w$.\label{thm:ohidepresref}
\end{theorem}

\begin{theorem}[Input Hiding Preserves Refinement]
If $A, B$ are both sink interfaces and $A \modabs B$, then $\ihide(w, A) \modabs \ihide(w, B)$ for any variable $w$. \label{thm:ihide}
\end{theorem}

Proofs for Theorems \ref{thm:comppresref} and \ref{thm:ohidepresref} are provided in \cite{tripakis2011theory}. \Cref{thm:ihide}'s proof is simple and is omitted.
One can think of using interface composition and variable hiding to horizontally (with respect to the refinement order) navigate the space of all interfaces. The synthesis pipeline encodes one navigated path and monotonicity of these operators yields guarantees about the path's end point. Composite operators such as $\CPRE(\cdot)$ chain together multiple incremental steps. Furthermore since the composition of monotone operators is itself a monotone operator, any composite constructed from these parts is also monotone.
%Each interface signature has an associated partial order of interfaces. These operations move across domains and preserve the ordering.
In contrast, the coarsening and refinement operators introduced later in \Cref{def:coarsen} and \ref{def:refinement} respectively are used to move vertically and construct abstractions. The ``direction" of new composite operators can easily be established through simple reasoning about the cumulative directions of their constituent operators.

%The next two sections show the coarsening and refinement operations, which are instead used to move vertically along the refinement order to construct abstractions.

%\begin{corollary}
%The winning region in controller synthesis yields an underapproximation of the winning set. 
%\end{corollary}
%\begin{figure}
%\caption{Composition and both hiding operations preserve the refinement relation across domains.}
%\end{figure}

\subsection{Dynamically Coarsening Interfaces}

%An interface can be quantized by composing it in series with a special quantizer interface.
In practice, the sequence of interfaces $Z_i$ generated during synthesis grows in complexity. This occurs even if the dynamics $F$ and the target/safe sets have compact representations (i.e., fewer nodes if using BDDs). Coarsening $F$ and $Z_i$ combats this growth in complexity by effectively reducing the amount of information sent between iterations of the fixed point procedure.

Spatial discretization or \textit{coarsening} is achieved by use of a quantizer interface that implicitly aggregates points in a space into a partition or cover.
\begin{definition} \label{def:relquant}
A quantizer $Q(i,o)$ is any interface that abstracts the identity interface $(i == o)$ associated with the signature $(i,o)$.
\end{definition}

Quantizers decrease the complexity of the system representation and make synthesis more computationally tractable.
A coarsening operator abstracts an interface by connecting it in series with a quantizer. Coarsening reduces the number of non-blocking inputs and increases the output non-determinism.

%Informally, a quantizer $Q(\abs i, i)$ is an interface where the input-output domains match $\dom{\abs i} \sequiv \dom{i}$ and $Q$ ``abstracts" an identity interface.
%\todo{Can be interpreted as mapping to a power set, but a finite collection.}
%In the appendix, \Cref{sec:dynamicprecision} shows one instance of a quantizer that maps from $[0,1] \subset \mathbb{R}$ to sets in a finite cover, and a generalization to higher dimensions.

\begin{definition}[Input/Output Coarsening] \label{def:coarsen}
Given an interface $F(i,o)$ and input quantizer $\quant{}(\abs i, i)$, input coarsening yields an interface with signature $(\abs i, o)$.
\begin{align}
\incoarsen(F, Q(\abs i, i)) &= \ohide(i, \comp(Q(\abs i, i), F) ) \label{eqn:icoursen}
\end{align}
Similarly, given an output quantizer $\quant{}(o, \abs o)$, output coarsening yields an interface with signature $(i, \abs o)$.
\begin{align}
\outcoarsen(F, Q(o, \abs o)) &= \ohide(o, \comp( F, Q(o, \abs o))) \label{eqn:ocoursen}
\end{align}
\end{definition}
\begin{figure}[!t]
\centering
\includegraphics[width=.32\columnwidth]{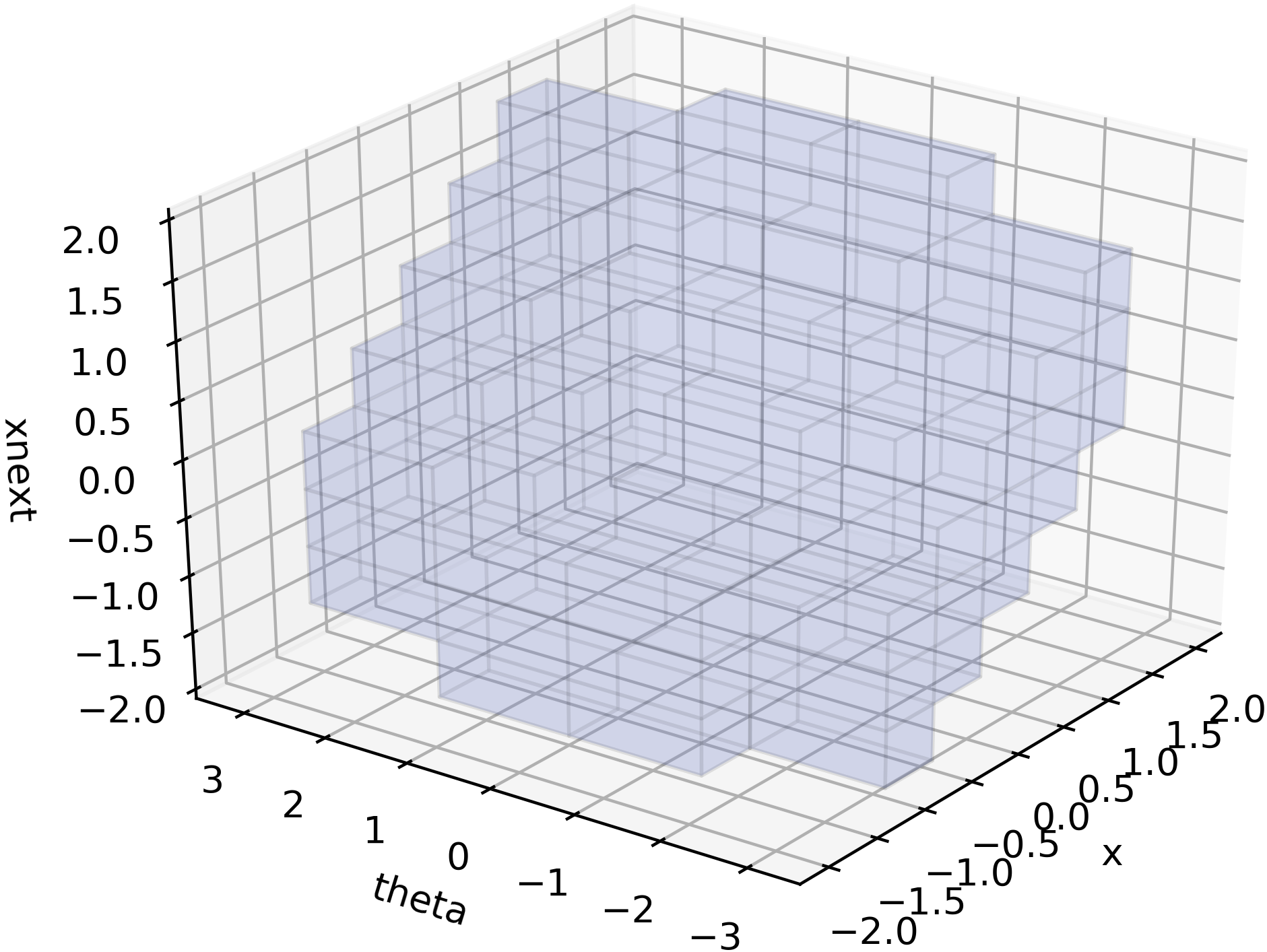}
\includegraphics[width=.32\columnwidth]{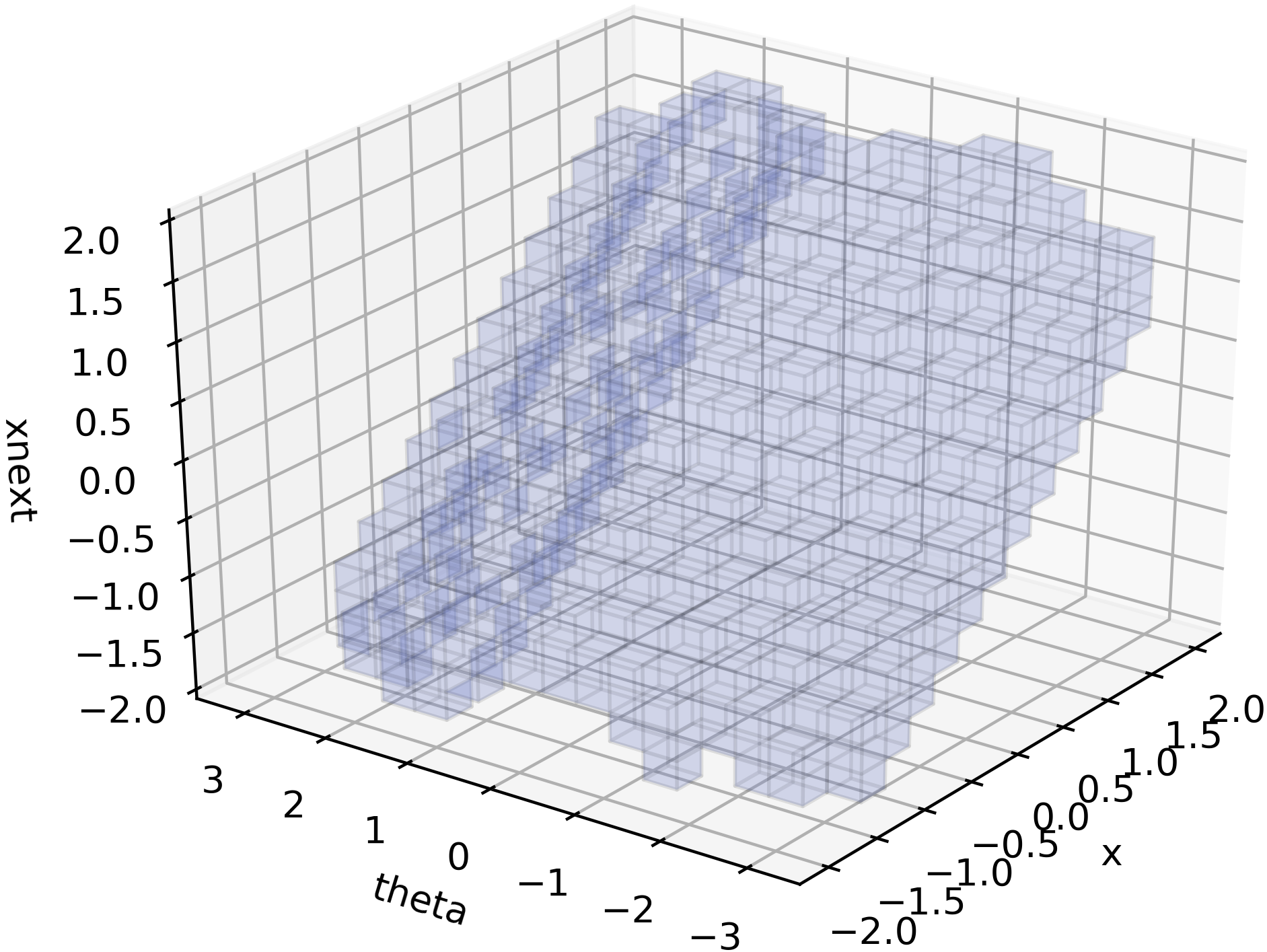}
\includegraphics[width=.32\columnwidth]{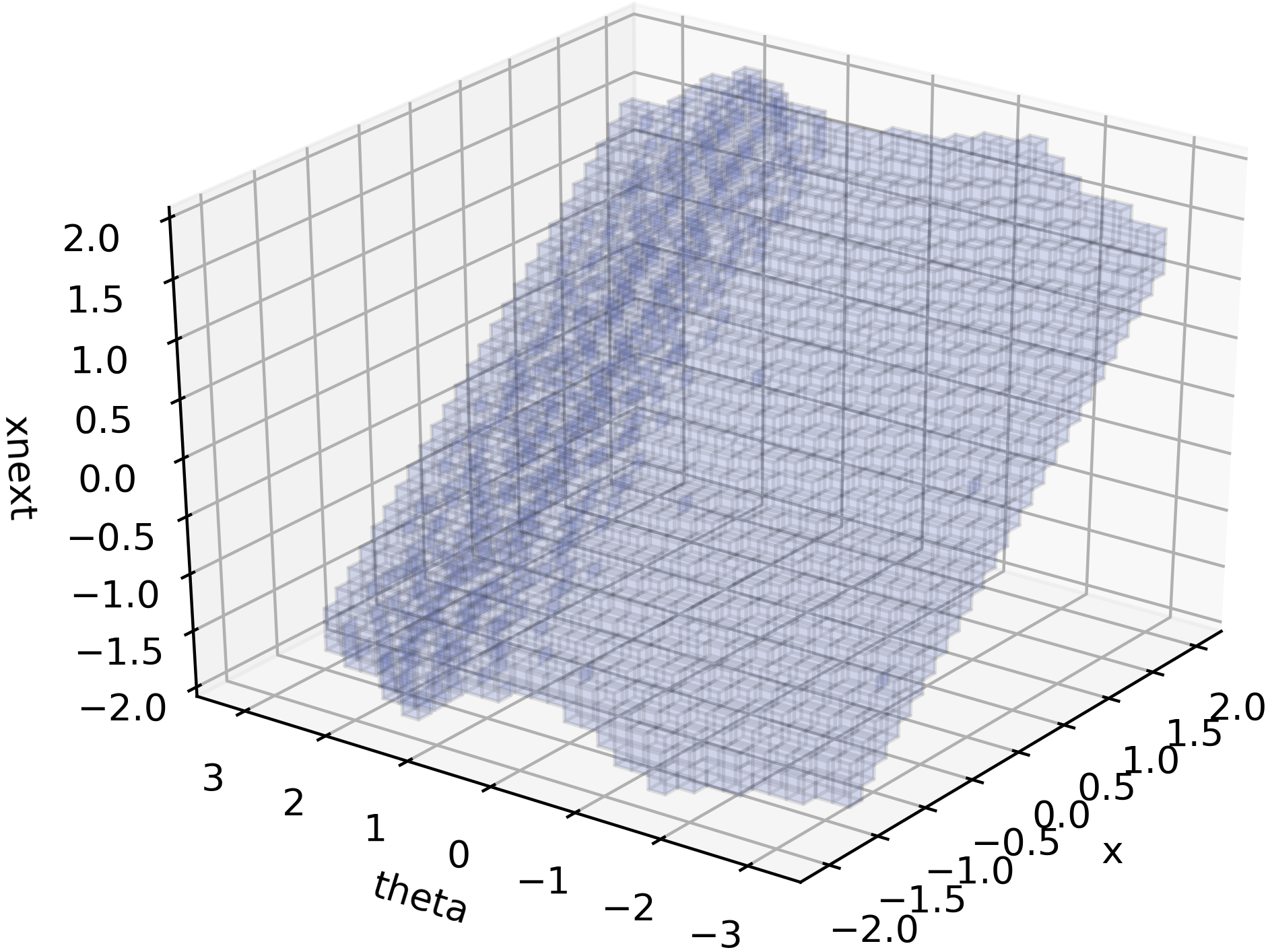}
\caption{\small Coarsening of the $F_x$ interface to $2^3, 2^4$ and $2^5$ bins along each dimension for a fixed $v$ assignment. %The inserted quantizer can be interpreted as coarsening either $F_x$'s output, $Z$'s input, or both. 
Interfaces are coarsened within milliseconds for BDDs but the runtime depends on the finite abstraction's data structure representation.
} \label{fig:coarsesynth}
\end{figure}
\Cref{fig:coarsesynth} depicts how coarsening reduces the information required to encode a finite interface. 
\ifcav
It leverages a variable precision quantizer, whose implementation is described in the extended version at \cite{redax}.
\else
\Cref{sec:dynamicprecision} in the appendix shows how the dynamic precision quantizer in \cite{redax} is implemented.
\fi

The corollary below shows that quantizers can be seamlessly integrated into the synthesis pipeline while preserving the refinement order. It readily follows from \Cref{thm:comppresref}, \Cref{thm:ohidepresref}, and the quantizer definition.
\begin{corollary}
Input and output coarsening operations\pref{eqn:icoursen} and \pref{eqn:ocoursen} are monotone operations with respect to the interface refinement order $\modabs$.
\end{corollary}

%\subsection{Coarsening During Synthesis}
\newcommand{\reach}{\code{reach}}
%, continuously applying $\reach(\cdot)$ yields a sequence of BDD interfaces that grows in complexity.
% BDDs are implemented as graphs and their complexity is measured in the number of nodes. For the lunar lander problem, many computers would run out of memory before the game solver reaches a fixed point.

It is difficult to know a priori where a specific problem instance lies along the spectrum between mathematical precision and computational efficiency. It is then desirable to coarsen dynamically in response to runtime conditions rather than statically beforehand.
%To maintain tractability, interfaces for both the winning region and dynamics are coarsened whenever the winning region becomes too complex.
\begin{figure}[!t]
\centering 
\includegraphics[width=.65\columnwidth]{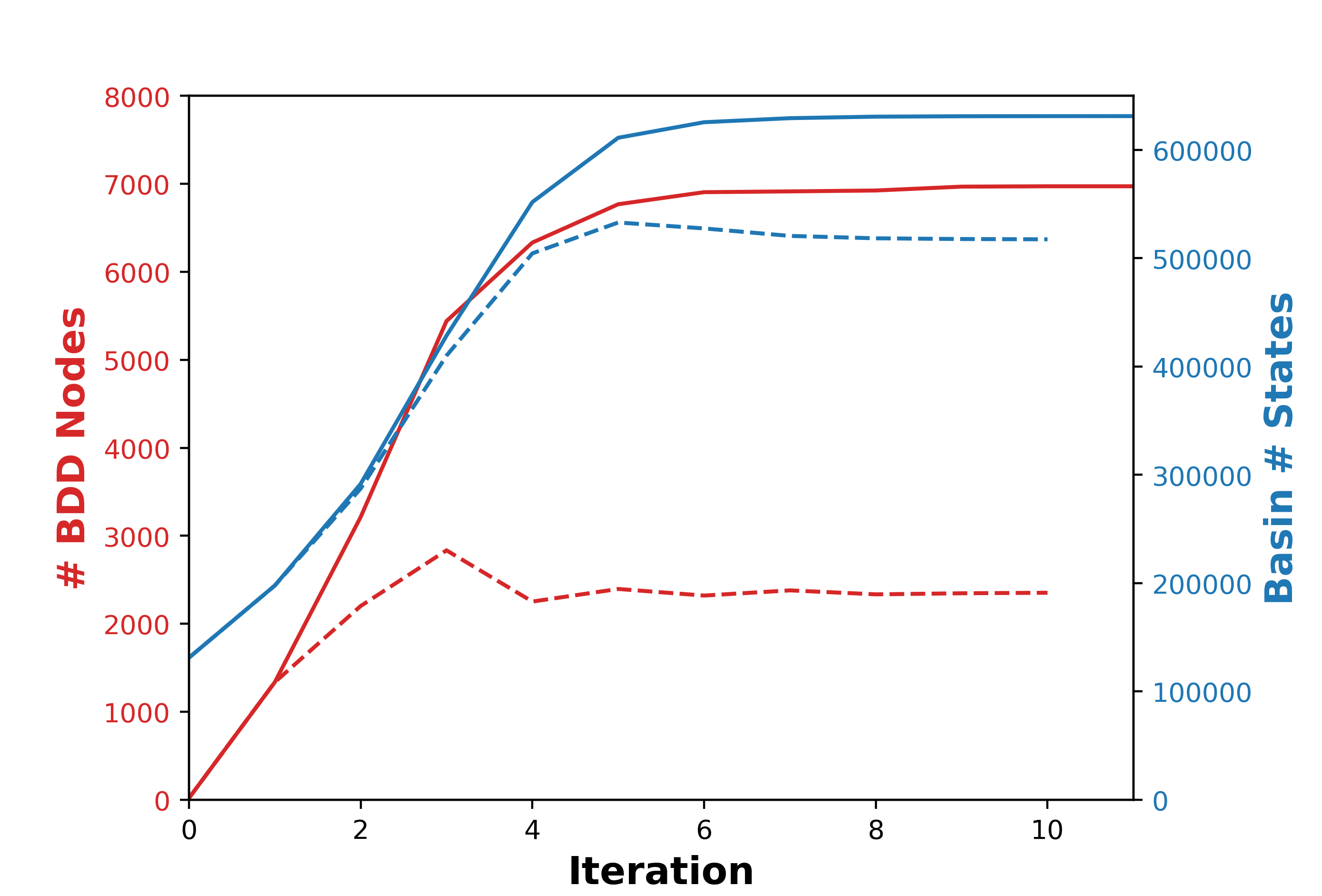}
\caption{\small Number of BDD nodes (red) and number of states in reach basin (blue) with respect to the reach game iteration with a greedy quantization. The solid lines result from the unmodified game with no coarsening heuristic. The dashed lines result from greedy coarsening whenever the winning region exceeds 3000 BDD nodes.} \label{fig:coarsenheuriters}
\end{figure}
Coarsening heuristics for reach games include:
\begin{itemize}[leftmargin=3.5mm]
\item \textit{Downsampling with progress} \cite{Hsu:2018:MAC:3178126.3178143}:
Initially use coarser system dynamics to rapidly identify a coarse reach basin.
Finer dynamics are used to construct a more granular set whenever the coarse iteration ``stalls".
%Whenever the game iteration is hasn't a fixed point, a coarser representation of the dynamics is used.
In \cite{Hsu:2018:MAC:3178126.3178143} only the $Z_i$ are coarsened during synthesis. We enable the dynamics $F$ to be as well.

%Coarse interfaces of the system dynamics are initially used to rapidly identify 
%Coarsening is done to reduce the number of fixed point iterations.
%The multi-layered approach precomputes abstractions of multiple granularities whereas we do not. Multi-layered synthesis starts with the coarsest dynamics and monitors progress (or lack thereof) in the fixed point calculation to decide when to switch amongst the levels whereas our approach only monitors usage of computational resources. Multi-layered coarsening also occurs along each state dimension uniformly and simultaneously. \todo{If sufficient progress made on finer models then go coarser}

\item \textit{Greedy quantization}:
Selectively coarsening along certain dimensions by checking at runtime which dimension, when coarsened, would cause $Z_i$ to shrink the least. This reward function 
can be leveraged in practice because coarsening is computationally cheaper than composition.
For BDDs, the winning region can be coarsened until the number of nodes reduces below a desired threshold. \Cref{fig:coarsenheuriters} shows this heuristic being applied to reduce memory usage at the expense of answer fidelity. A fixed point is not guaranteed as long as quantizers can be dynamically inserted into the synthesis pipeline, but is once quantizers are always inserted at a fixed precision.
\end{itemize}

The most common quantizer in the literature never blocks and only increases non-determinism (such quantizers are called ``strict" in \cite{gunther2}\cite{Rungger2016}). If a quantizer is interpreted as a partition or cover, this requirement means that the union must be equal to an entire space. \Cref{def:relquant} relaxes that requirement so the union can be a subset instead. It also hints at other variants such as interfaces that don't increase output non-determinism but instead block for more inputs.

%\section{Quantized Interfaces} \label{sec:quantinter}

\tikzset{
block/.style = {draw, fill=white, rectangle, minimum height=3em, minimum width=3em},
tmp/.style  = {coordinate}, 
sum/.style= {draw, fill=white, circle, node distance=1cm},
input/.style = {coordinate},
output/.style= {coordinate},
pinstyle/.style = {pin edge={to-,thin,black}
}
}

\section{Refining System Dynamics} \label{sec:abs}

Shared refinement \cite{tripakis2011theory} is an operation that takes two interfaces and merges them into a single interface. In contrast to coarsening, it makes interfaces more precise.
Many tools construct system abstractions by starting from the universal abstraction $\false$, then iteratively refining it with a collection of smaller interfaces that represent input-output samples.
This approach is especially useful if the canonical concrete system is a black box function, Simulink model, or source code file. These representations do not readily lend themselves to the predicate operations or be coarsened directly.
We will describe later how other tools implement a restrictive form of refinement that introduces unnecessary dependencies.

%There are many data structures that could be used to represent finite domain predicates including lookup tables, n-dimensional arrays with binary values, trees, bitmaps, and boolean circuits.
%Directly translating Simulink models (for instance) into any of the above data structures is non-trivial.

Interfaces can be successfully merged whenever they do not contain contradictory information.
The shared refinability condition below formalizes when such a contradiction does not exist.
\begin{definition}[Shared Refinability \cite{tripakis2011theory}] \label{def:refinable}
Interfaces $F_1(i,o)$ and $F_2(i,o)$ with identical signatures are shared refinable if
\begin{align}
\left( \NB{F_1} \AND \NB{F_2} \right) \implies \equant{o} (F_1 \AND F_2)
\label{eqn:refinability}
\end{align}
\end{definition}
For any inputs that do not block for all interfaces, the corresponding output sets must have a non-empty intersection.
%\todo{This condition is satisfied for forward reachable set overapproximations because overlapping initial states set must result in overlapping next state sets.}
If multiple shared refinable interfaces, then they can be combined into a single one that encapsulates all of their information.
\begin{definition}[Shared Refinement Operation \cite{tripakis2011theory}] \label{def:refinement}
The shared refinement operation combines two shared refinable interfaces $F_1$ and $F_2$, yielding a new identical signature interface corresponding to the predicate
\begin{align}
\code{refine}(F_1, F_2) = (\NB{F_1} \OR \NB{F_2}) \AND (\NB{F_1} \implies F_1) \AND (\NB{F_2} \implies F_2). \label{eqn:refine}
\end{align}
%The shared refinement operation combines a collection of interfaces $M_1, \ldots, M_N$ that is pairwise shared refinable and yields a new interface:
%\begin{align}
%\code{refine}(M_1, \ldots, M_N) =
%\left (\bigvee_{j=1}^N \NB{M_j}\right) \AND \bigwedge_{j=1}^N \left(\NB{M_j} \implies M_j \right) \label{eqn:refine}
%\end{align}
\end{definition}
The left term expands the set of accepted inputs. The right term signifies that if an input was accepted by multiple interfaces, the output must be consistent with each of them. 
The shared refinement operation reduces to disjunction for sink interfaces and to conjunction for source interfaces.

Shared refinement's effect is to move up the refinement order by combining interfaces.
Given a collection of shared refinable interfaces, the shared refinement operation yields the least upper bound with respect to the refinement partial order in \Cref{def:relabs}.
Violation of \pref{eqn:refinability} can be detected if the interfaces fed into $\code{refine}(\cdot)$ are not abstractions of the resulting interface.
%This is a useful check for ensuring that reach set overapproximations are computed correctly.
%Failure of this condition was a reliable indicator that there was a bug when setting up \Cref{sec:examples}'s examples.

%\begin{theorem}[Shared Refinement Yields a Least Upper Bound \cite{tripakis2011theory}] \label{thm:refinementlub}
%
%\end{theorem}

\subsection{Constructing Finite Interfaces through Shared Refinement}

%\todo{Sometimes you can't compute overapproximations or simulate, can use data from a file.}

A common method to construct finite abstractions is through simulation and overapproximation of forward reachable sets. This technique appears in tools such as PESSOA \cite{mazo2010pessoa}, SCOTS \cite{Rungger2016}, MASCOT \cite{Hsu:2018:MAC:3178126.3178143}, ROCS \cite{rocs} and ARCS \cite{novelbddencoding}.
By covering a sufficiently large portion of the interface input space, one can construct larger composite interfaces from smaller ones via shared refinement.

\begin{figure}
\centering
\includegraphics[width=.5\columnwidth]{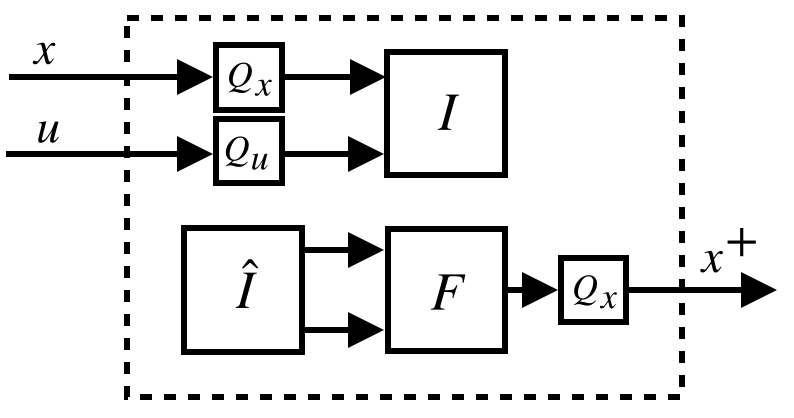}
\includegraphics[width=.42\columnwidth, height = 1.2in]{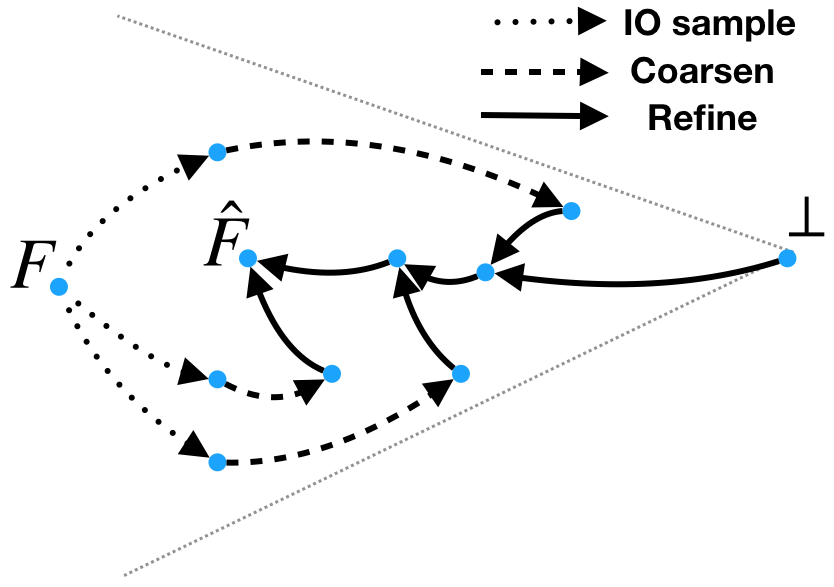}
\caption{\small (Left) Result of sample and coarsen operations for control system interface $F(x \cup u, x^+)$. The $I$ and $\hat{I}$ interfaces encode the same predicate, but play different roles as sink and source. (Right) Visualization of finite abstraction as traversing the refinement partial order. Nodes represent interfaces and edges signify data dependencies for interface manipulation operators. Multiple refine edges point to a single node because refinement combines multiple interfaces. Input-output (IO) sample and coarsening are unary operations so the resulting nodes only have one incoming edge. The concrete interface $F$ refines all others, and the final result is an abstraction $\abs F$. %\Cref{alg:disjunctive} (left) yields a worse abstraction than \Cref{alg:refine_abs} (right) because the initial samples contain less information.
} \label{fig:samplequant}
\end{figure}
Smaller interfaces are constructed by sampling regions of the input space and constructing an input-output pair.
In \Cref{fig:samplequant}'s left half, a sink interface $I(x \cup u, \emptyset)$ acts as a filter. The source interface $\hat{I}(\emptyset, x \cup u)$ composed with $F(x \cup u, x^+)$ prunes any information that is outside the relevant input region.
The original interface refines any sampled interface. To make samples \textit{finite}, interface inputs and outputs are coarsened. An individual sampled abstraction is not useful for synthesis because it is restricted to a local portion of the interface input space. After sampling many finite interfaces are merged through shared refinement. 
The assumption $\abs{I}_i \implies \NB{F}$ encodes that the dynamics won't raise an error when simulated and is often made implicitly.
\Cref{fig:samplequant}'s right half depicts the sample, coarsen, and refine operations as methods to vertically traverse the interface refinement order.

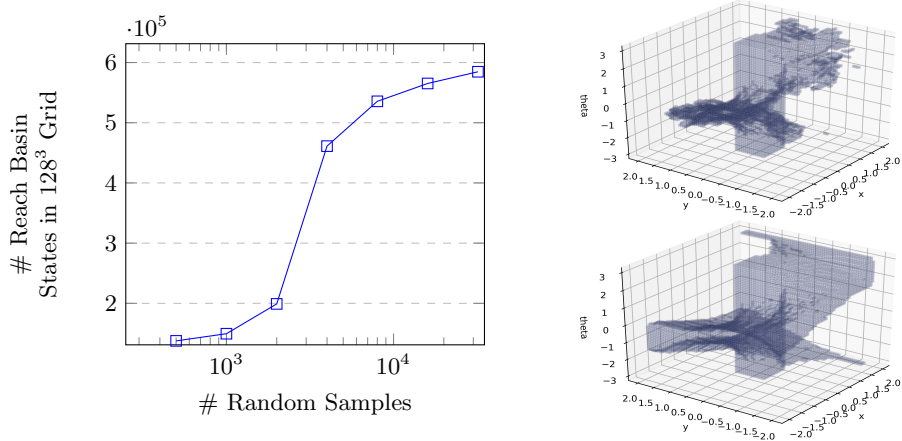
\begin{figure}
\centering
\begin{minipage}{.62\columnwidth}
\begin{tikzpicture}
\begin{semilogxaxis}[
    xlabel={ \# Random Samples},
    ylabel style={align=center},
    ylabel={\# Reach Basin\\ States in $128^3$ Grid},
    xmin=250, xmax=35000,
    ymin=131072, ymax=631272,
    xtick={0,100,1000,10000,100000},
    ytick={0,100000,200000,300000,400000,500000,600000},
    ymajorgrids=true,
    grid style=dashed,
    width=2.5in,
    height=2.2in
    %    legend pos=north west,
]
 
\addplot[
    color=blue,
    mark=square,
    ]
    coordinates {
    (500,137630)(1000,149540)(2000,199040)(4000,461150)(8000,535580)(16000,565230)(32000,584670)
    };   
\end{semilogxaxis}
\end{tikzpicture}
\end{minipage}
\begin{minipage}{.35\columnwidth}
\includegraphics[width=\columnwidth,  height=1.15in]{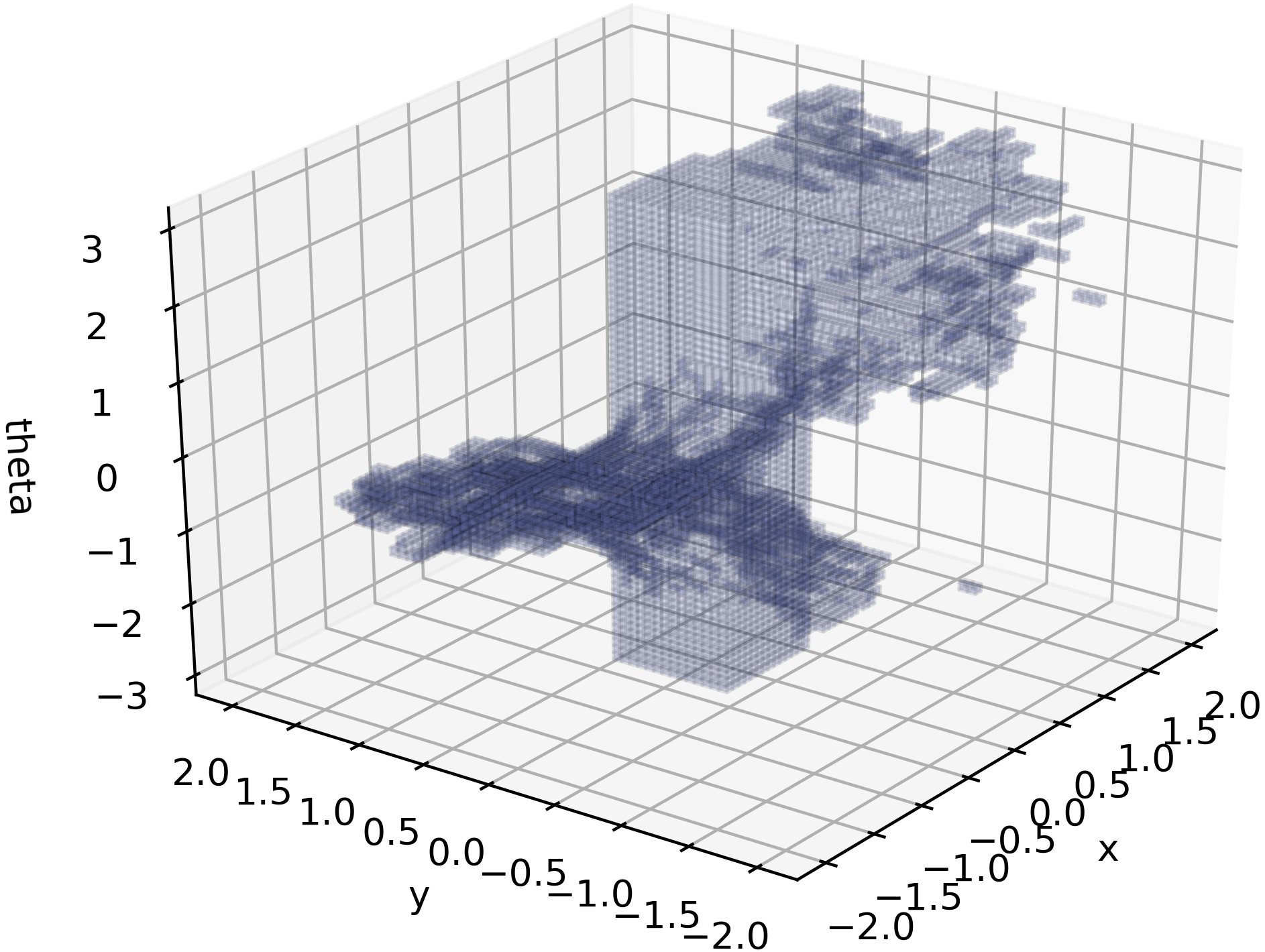}\\
\includegraphics[width=\columnwidth,  height=1.15in]{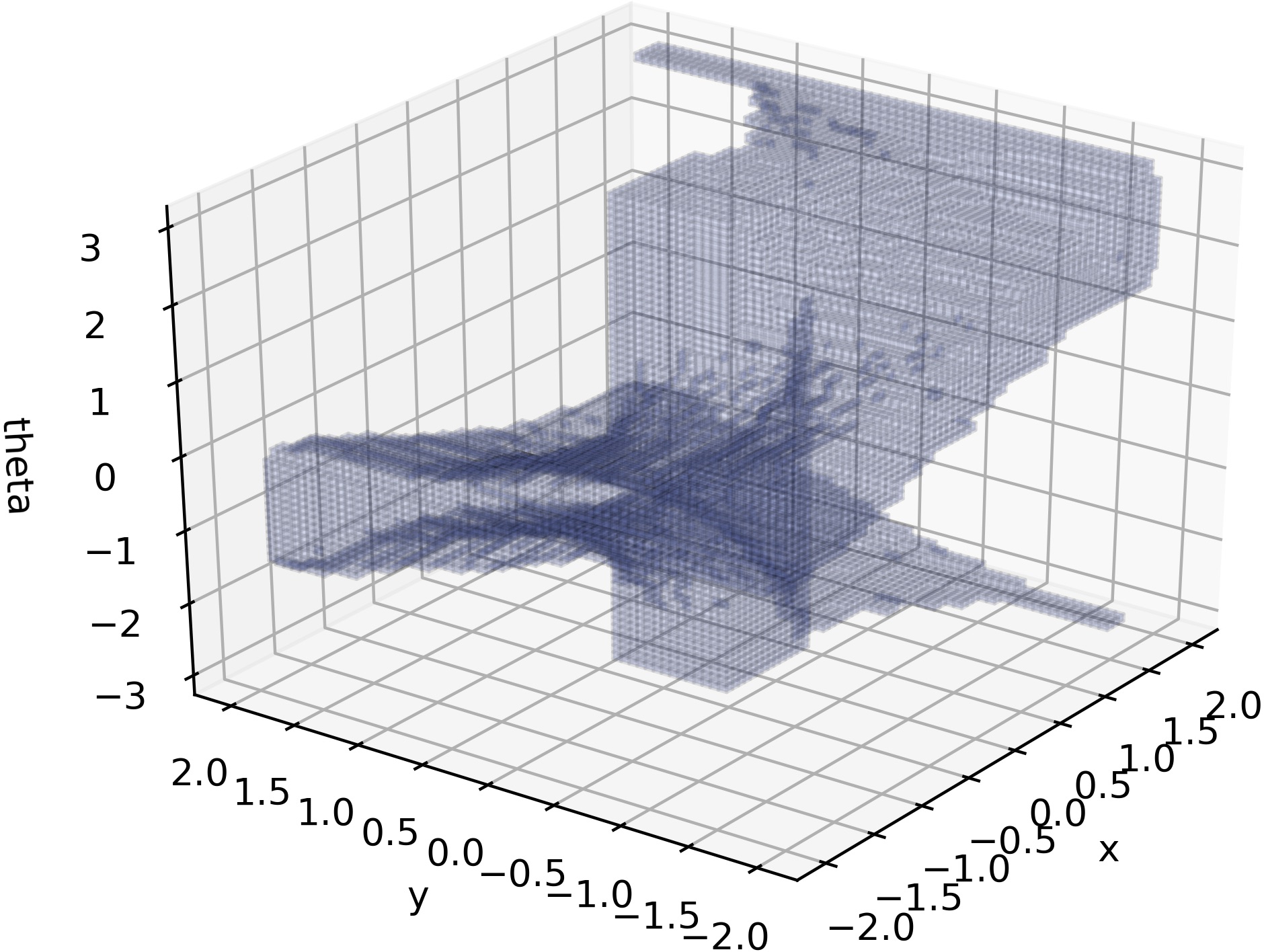}
\end{minipage}
\caption{\small The number of states in the computed reach basin grows with the number of  random samples. The vertical axis is lower bounded by the number of states in the target $131k$ and upper bounded by $631k$, the number of states using an exhaustive traversal. Naive implementations of the exhaustive traversal would require $12$ million samples. The right shows basins for 3000 (top) and 6000 samples (bottom).} \label{fig:growbasinwithsamples}
\end{figure}

Critically, $\code{refine}(\cdot)$ can be called within the synthesis pipeline and does not assume that the sampled interfaces are disjoint.
\Cref{fig:growbasinwithsamples} shows the results from refining the dynamics with a collection of state-control hyper-rectangles that are randomly generated via uniformly sampling their widths and offsets along each dimension. These hyper-rectangles may overlap. If the same collection of hyper-rectangles were used in MASCOT, SCOTS, ARCS, or ROCS then this would yield a much more conservative abstraction of the dynamics because their implementations are not robust to overlapping or misaligned samples.
PESSOA and SCOTS circumvent this issue altogether by enforcing disjointness with an exhaustive traversal of the state-control space, at the cost of unnecessarily coupling the refinement and sampling procedures. 
\ifcav
The lunar lander in the extended version \cite{redax} embraces overlapping and uses two mis-aligned grids to construct a grid partition with $p^N$ elements with only $p^N (\frac{1}{2})^{N-1}$ samples (where $p$ is the number of bins along each dimension and $N$ is the interface input dimension). This technique introduces a small degree of conservatism but its computational savings typically outweigh this cost.
\else
The lunar lander contained in \Cref{sec:lunar} of the appendix embraces overlapping and uses two mis-aligned grids to construct a grid partition with $p^N$ elements with only $p^N (\frac{1}{2})^{N-1}$ samples (where $p$ is the number of bins along each dimension and $N$ is the interface input dimension). This technique introduces a small degree of conservatism but its computational savings typically outweigh this cost.
\fi
%ROCS \cite{rocs} and ARCS \cite{novelbddencoding} incorporate a restrictive/specialized form of $\code{refine}(\cdot)$ into the synthesis pipeline to obtain more precise and useful results.

%Ill-chosen quantization parameters can cause the synthesis step to either yield trivial results or run out of memory. In practice, users must tune hyperparameters until they identify the proper balance between granularity and computational feasibility.
%Additional interfaces can be incorporated into the abstraction at any time during the synthesis procedure; this is useful if the result that is too conservative. % via the shared refinement step in \Cref{eqn:binrefine}.
%Their sampling methods are enumerative. That is, the number of discrete states is equivalent to the number of samples.
%One could even refine interfaces with random boxes generated by uniformly sampling their widths and offsets along each dimension. Unlike similar procedures in ROCS and ARCS, the offsets and widths do \textit{not} need to 
%\Cref{fig:growbasinwithsamples} shows how the reach basin grows with respect to the number of samples.

\section{Decomposed Control Predecessor} \label{sec:decomppre}

%Iterative synthesis that is done component-wise.

%They can exhibit additional internal structure because they are constructed from smaller internal components. 
%The Dubin's vehicle contains three interfaces encoding continuous relations:
%\begin{align}
%p_x^+ &== p_x + v \cos(\theta)\\
%p_y^+ &== p_y + v \sin(\theta) \\
%\theta^+ &== \theta + \frac{v}{L} \sin(\omega) 
%\end{align}

%The controlled predecessor from \Cref{sec:core} requires a monolithic control system interface.
%Constructing a symbolic representation of the monolithic interface $F$ by preemptively collapsing its components down via the interface composition operator $\comp(\cdot)$ may be computationally expensive.
%and does not take advantage of $F$'s decomposition into a collection of simpler interfaces.

%One simple observation is that
%transition relations over continuous spaces often are represented via component-wise updates along each state dimension. For instance, the Dubins vehicle is represented by the three interfaces $F_x, F_y$, and $F_\theta$.

A decomposed control predecessor is available whenever the system state space consists of a Cartesian product and the dynamics are decomposed component-wise such as $F_x, F_y$, and $F_\theta$ for the Dubins vehicle. This property is common for continuous control systems over Euclidean spaces.
While one may construct $F$ directly via the abstraction sampling approach, it is often intractable for larger dimensional systems. 
A more sophisticated approach abstracts the lower dimensional components $F_x, F_y$, and $F_\theta$ individually, computes $F = \comp(F_x, F_y, F_\theta)$, then feeds it to the monolithic $\CPRE(\cdot)$ from \Cref{prop:cpreop}. This section's approach is to avoid computing $F$ at all and decompose the monolithic $\CPRE(\cdot)$. It operates by breaking apart the term $\ohide(x^+, \comp(F,Z))$ in such a way that it respects the decomposition structure.
%and does not take advantage of $F$'s decomposition into a collection of simpler interfaces.
\begin{figure}[!t]
\centering
\includegraphics[width=.85\columnwidth]{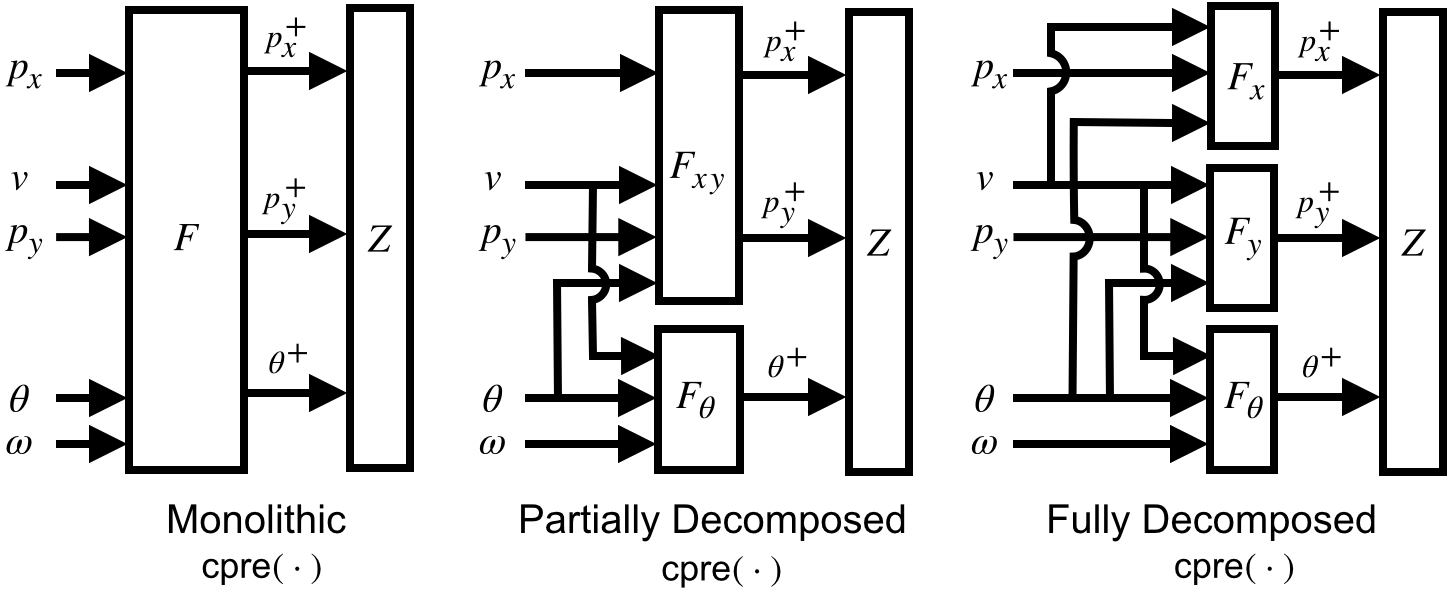}
\begin{tabular}{|c|c|c|}
\hline
\textbf{Decomposition} & \textbf{Parallel Compose}  & \textbf{Reach Game}\\
& \textbf{Runtime (s)} & \textbf{Runtime (s)}\\ \hline \hline
$F$ (Monolithic) & 0.56 & 103.09 \\ \hline
$F_{y\theta}, F_x$ (Partially Decomp.) & 0.02 & 28.31 \\ \hline
$F_{x\theta}, F_y$ (Partially Decomp.) & 0.01 & 28.71 \\ \hline
$F_{xy}, F_\theta$ (Partially Decomp.) & 0.06 & 10.61 \\ \hline
$F_x, F_y, F_\theta$ (Fully Decomp.) & n/a & 4.42 \\ \hline
\end{tabular}
\caption{\small A monolithic $\CPRE(\cdot)$ incurs unnecessary pre-processing and synthesis runtime costs for the Dubins vehicle reach game. Each variant of $\CPRE(\cdot)$ above composes the interfaces $F_x, F_y$ and $F_\theta$ in different permutations. For example, $F_{xy}$ represents $ \comp(F_x, F_y)$ and $F$ represents $\comp(F_x, F_y, F_\theta)$.
%A monolithic $\CPRE(\cdot)$ computes the monolithic dynamics $F$ first while a fully decomposed $\CPRE(\cdot)$ never constructs $F$.
} \label{fig:monopartifulldecomp}
\end{figure}
For the Dubins vehicle example $\ohide(x^+, \comp(F,Z))$ is replaced with
%\begin{align}
%\ohide(p_x^+, \comp(F_x, \ohide(p_y^+, \comp(F_y, \ohide(\theta^+, \comp(F_\theta, Z)))))) \label{eqn:dubpreZ}
%\end{align}
\[\ohide(p_x^+, \comp(F_x, \ohide(p_y^+, \comp(F_y, \ohide(\theta^+, \comp(F_\theta, Z))))))\]
yielding a sink interface with inputs $p_x, p_y, v, \theta$, and $\omega$.
%that avoids constructing $F$.
This representation and the original $\ohide(x^+, \comp(F,Z))$ are equivalent because $\comp(\cdot)$ is associative and interfaces do not share outputs $x^+ \sequiv \{p_x^+, p_y^+, \theta^+\}$.
%Computing \Cref{dubinsapprox} with a decomposed predecessor requires $4.146s$ versus $296.78s$ for the monolithic one.
\Cref{fig:monopartifulldecomp} shows multiple variants of $\CPRE(\cdot)$ and improved runtimes when one avoids preemptively constructing the monolithic interface.
The decomposed $\CPRE(\cdot)$ resembles techniques to exploit partitioned transition relations in symbolic model checking \cite{Burch:1991}.

No tools from \Cref{subsec:abcs} natively support decomposed control predecessors.
We've shown a decomposed abstraction for components composed in parallel but this can also be generalized to series composition to capture, for example, a system where multiple components have different temporal sampling periods.

\ifcav
%\todo{Heterogeneous grids in lunar lander} 
\else
%\todo{This decomposition is not only useful for synthesis but also abstraction. Each component can use a different precision quantizer which....... }
\fi

%\begin{figure}
%\centering
%\begin{subfigure}[t]{.39\columnwidth}
%\includegraphics[height=1.4in]{figs/dubinsdecomppre.png}
%\caption{\small A decomposed controlled predecessor for an interconnection containing interfaces $F_x$, $F_y$, and $F_\theta$. $F$ is equivalent to the parallel composition of those interfaces.} \label{fig:decomppre}
%\end{subfigure} \hspace{.4cm}
%\begin{subfigure}[t]{.55\columnwidth}
%\includegraphics[height=1.4in]{figs/quantinsynth.png}
%\caption{\small A decomposed control predecessor that incorporates a quantizer. Associativity of $\comp(\cdot)$ means $Q_x$ can be viewed as quantizing $Z$ , $F_x$, or both.} \end{subfigure}
%\caption{Two decomposed control predecessors.}
%\end{figure}%\input{examples}
%\input{reflection}
\section{Conclusion}

Tackling difficult control synthesis problems will require exploiting \textit{all} available structure in a system with tools that can \textit{flexibly adapt} to an individual problem's idiosyncrasies.
This paper lays a foundation for developing an extensible suite of interoperable techniques and demonstrates the potential computational gains in an application to controller synthesis with finite abstractions.
%%Tackling high dimensional control synthesis will require a variety of theoretical and algorithmic techniques and a method to combine them.
Adhering to a simple yet powerful set of well-understood primitives also constitutes a disciplined methodology for algorithm development, which is especially necessary if one wants to develop concurrent or distributed algorithms for synthesis.
%By adopting a small, core framework with relational interfaces it is easier for synthesis algorithms to dynamically reconfigure themselves at runtime and explore trade-offs between different requirements.
%Next steps will focus on safely leveraging 
%\todo{Library? Meta level. Mix match. Align with computing platforms.}
%While many existing operations were showcased as instances in this framework, we believe that  many additional ones that can readily be encoded such as specification-driven abstraction or abstractions over multiple time horizons. 
%
%Restricting one's self to a simple, yet powerful collection of well-understood primitives gives a safe and disciplined foundation for exploring future algorithmic modifications.

%%key contribution is to make it easier to explore the space of optimizations. 
%%The examples in this paper did not incorporate features like specification-driven abstraction or abstractions over multiple time horizons.
%

\ifcav
% No appendix in the cav version
\else
\section{Appendix}

%\subsection{Proof of \Cref{prop:cpreop}} \label{cprepropproof}
%Applying the definitions of $\comp$, $\ihide$, and $\ohide$ yields the expression 
%\begin{align}
%\equant{u}\equant{x^+}(F \AND Z \AND \uquant{x^+}(F \implies Z)))
%\end{align}
%One can safely move the $\equant{x^+}$ inside the parenthesis because $\uquant{x^+}(F \implies Z)$ is a predicate that is independent of $x^+$, yielding
%\begin{align}
%\equant{u}(\equant{x^+}(F \AND Z) \AND \uquant{x^+}(F \implies Z)))
%\end{align}
%To show equivalence of the above expressions with \Cref{eqn:origcpre}, we simply need to show equivalence of the following two predicates that depend on $x$ and $u$:
%\begin{align}
%\equant{x^+}(F \AND Z) \AND \uquant{x^+}(F \implies Z) \label{eqn:xu1}\\
%\equant{x^+} F \AND \uquant{x^+}(F \implies Z). \label{eqn:xu2}
%\end{align}
%It is easy to see that \pref{eqn:xu1} implies \pref{eqn:xu2} because $\equant{x^+}(F\AND Z)$ implies $\equant{x^+}F$. To show the reverse, suppose \pref{eqn:xu2} is satisfied for a pair $x$ and $u$. Any $x^+$ chosen to satisfy $\equant{x^+}F$ must also satisfy the constraint imposed by the sink $Z$. Otherwise, the clause $\uquant{x^+}(F \implies Z)$ would be violated, contradicting satisfaction of \pref{eqn:xu2}. Therefore, \pref{eqn:xu1} and \pref{eqn:xu2} are equivalent which completes the proof.

\subsection{Dynamic Precision Quantization}
\label{sec:dynamicprecision}

%We show a particular quantizer that is geared towards dynamic or variable precision abstractions such as those that appear in tools ROCS, ARCS, and MASCOT. Afterwards, we show how the quantizer can be used to coarsen interfaces via series composition.
%For clarity, our quantizer will build off a one-dimensional quantizer that translates a real number within a compact interval into a finite vector of bits. \Cref{rem:quantizer} will briefly explain how to construct quantizers for higher dimension spaces and unbounded domains.

Certain data structures like trees or binary decision diagrams are natural candidates for encoding hierarchical decompositions of spaces.
A sequence of bits can be used to traverse that data structure to arrive at a subset of that space. More bits allow one to specify finer granularity sets. 
%\todo{This type of work is easiest to implement in a tree or decision diagram data structure. It is difficult with lookup tables or arrays because of indexing.}
%An an illustrative example, consider an input-output relation $M(i,o)$ where $\dom{i} \equiv \dom{o} \equiv [0,1]$. 
%Any value in $[0,1]$ can be associated with an infinite sequence of bits.
As an illustrative example, consider an interval $[0,1]$ that is to be covered. Let $N$ represent a number of bits to construct a cover of $[0,1]$, and assign a unique bit sequence to identify each set in the cover.
%Let $o = o_1 \cup o_1 \cup \ldots \cup o_{N}$ represent an ordered set of bits with length $N$.
\Cref{fig:hierarchical} depicts for varying $N$ how an interval can be covered by a collection of small intervals, while minimizing overlaps. A greater $N$ signifies that the cover can be constructed with finer intervals. A bit vector of length $N$ can be used to specify an interval of width $2^{-N}$. If $N = \infty$, then any value in $[0,1]$ can be encoded via the infinite weighted sum $\sum_{k=1}^\infty 2^{-k} o_k$. For $N=0$, the interval is $[0,1]$ itself. \Cref{fig:hierarchical} implicitly appends don't care terms $-$ to the end of bit vectors. This allows finite length bit vectors (intervals) to be identical to the disjunction (union) of longer bit vectors (smaller intervals) with a common prefix.

%For fixed $N$, this collection of intervals can be generated by applying the formula $\left[ \sum_{k=1}^{N} 2^{-k} o_k, 2^{-N} + \sum_{k=1}^{N} 2^{-k} o_k \right]$ for all $2^N$ combinations of values of $o = o_1 \cup o_1 \cup \ldots \cup o_{N}$ and interpreting the bits as integer values $o_k \in \{0,1\}$.
%If $N = \infty$, then the interval collapses to a single real value. 

%Suppose a floating point value on a computer is associated with a finite sequence of bits \footnote{Finiteness allows us to sidestep the fact that a real number can have multiple valid encodings as infinite sequences of bits.}. This sequence would encode an interval which contains that point, with the sequence length determining the interval's width.

A quantizer acts by truncating a bit vector to a specified finite number of bits and outputting the result. The effect of truncating bits is to implicitly widen or coarsen that interval. A quantizer $\quant{N}(o, \abs o)$ that only retains the first $N$ bits satisfies the following formula and is depicted in \Cref{fig:hierarchical}
\begin{align}
\quant{N}(o, \abs o) &= \quant{N}( o_1 \cup \ldots, \abs o_1 \cup \ldots) = \bigwedge_{k=1}^{N} \left( o_k == \abs o_k \right). \label{eqn:bitquant}
\end{align}

\begin{figure}
\centering
\newcommand{\dc}{-}
\begin{subfigure}{0.45\textwidth}
\begin{tabular}{|c|| c c c c }
  \hline
 $N=0$ & \multicolumn{4}{|c|}{$\dc\dc\dc$}\\ \hline
 $N=1$ & \multicolumn{2}{|c|}{$\false \dc \dc$} & \multicolumn{2}{c|}{$\true \dc \dc$} \\ \hline
  $N=2$&\multicolumn{1}{|c|}{$\false \false \dc$} 
  & \multicolumn{1}{c|}{$\false \true \dc$} 
  & \multicolumn{1}{c|}{$\true \false \dc$} 
  &\multicolumn{1}{c|}{$\true \true \dc$}  \\ \hline
%  $N=3$ &  \multicolumn{1}{|c|}{$\false \false \false$} 
%  & \multicolumn{1}{c|}{$\false\false \true$} 
%  & \multicolumn{1}{c|}{$\false\true \false$} 
%  &\multicolumn{1}{c|}{$\false\true \true$}
%  & \multicolumn{1}{|c|}{$\true\false \false$} 
%  & \multicolumn{1}{c|}{$\true\false \true$} 
%  & \multicolumn{1}{c|}{$\true\true \false$} 
%  &\multicolumn{1}{c|}{$\true\true \true$}\\ \hline
  $N > 2$ & \multicolumn{4}{|c|}{\vdots}\\ \hline
\end{tabular}
\end{subfigure}
\begin{subfigure}{0.45\textwidth}
\includegraphics[width=.6\columnwidth]{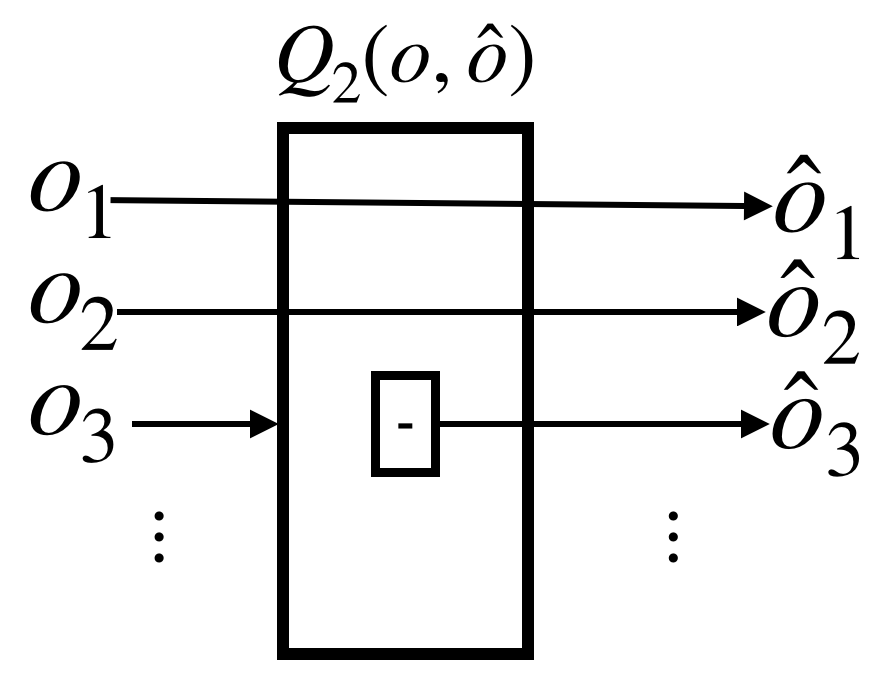}
\end{subfigure}
\caption{\small (Left) Varying length $N$ of a bit vector leads to different granularity covers of an interval. The don't care term ``$-$" signifies that the value of a certain bit does not matter. Bit vectors of finite length are implicitly appended with an infinite sequence of don't cares. (Right) A bit quantizer in \Cref{eqn:bitquant} with $N=2$. Higher significant bits are outputted but are replaced with a don't care term ``$-$" that can non-deterministically be true or false.} \label{fig:hierarchical}
\end{figure}

%\begin{figure} 
%\centering
%
%\caption{\small } \label{fig:bitquant}
%\end{figure}

%There are two visual interpretations of $\quant{N}(o, \abs o)$ as an interface depicted in \Cref{fig:bitquant} with $N=2$. In the first interpretation, the quantizer doesn't output any higher significant bits at all. With this interpretation the quantizer is a function where the output domain has a smaller cardinality than the input domain, as shown in in \Cref{fig:bitquant} which depicts $\quant{2}(o_1 \cup o_2 \cup \ldots, \abs o_1 \cup \abs o_2 )$.
The quantizer implicitly receives and emits infinite bit sequences but only depends on the most significant bits. It ignores higher precision input bits and non-deterministically assigns values to higher precision output bits. The input and output domains have identical cardinality but the quantizer is not a function due to the non-deterministic output.
%\begin{align}
%\quant{o}(o_0o_1\ldots, \abs o_0 \abs o_1 \ldots) &= \bigwedge_{k=0}^M \left( o_k == \abs o_k \right) \AND \true \\
%&= \bigwedge_{k=0}^M \left( o_k == \abs o_k \right) \AND \bigwedge_{k=M+1}^\infty (\neg o_k \OR o_k) \nonumber 
%\end{align}
It is easy to see that for two quantizers $\quant{N}, \quant{M}$ with precisions $N, M \in \mathbb{N} \cup \{\infty\}$ and $N \leq M $ the relation $\quant{N} \modabs \quant{M}$ holds. 
A collection of quantizers composed in series also yields a single quantizer with the minimum precision. 

Applying the quantizer definition above to the output coarsening and input coarsening operations yield insightful results.

\begin{example}[Output Quantizer]
Consider an interface $F(i, o)$ where $o = o_1 \cup o_2 \cup \ldots$ is a composite variable representing an infinite bit vector. Let $o_{msb} = o_1 \cup \ldots \cup o_N$ be the most significant bits and $o_{lsb} = o_{N+1} \cup \ldots$ be the least significant bits. 
%For any pair $(i, \abs o_{msb})$ that satisfies the interface.... $F(i, o_{msb}$ 
\begin{align}
&\outcoarsen(M, Q(o, \abs o))\\
&= \ohide(o, \comp( M, Q(o, \abs o)))\\
&= \equant o \left( \bigwedge_{k=1}^{N} \left( o_k == \abs{o_k} \right) \AND F \AND \uquant{o} \left(F  \implies \equant{\abs{o}} \left(\bigwedge_{k=1}^{N} \left( o_k == \abs{o_k} \right)\right) \right) \right)\\
&= \equant o \left( \bigwedge_{k=1}^{N} \left( o_k == \abs o_k \right) \AND F \AND \true \right)\\
&= \equant o_{msb} \left( \bigwedge_{k=1}^{N} \left( o_k == \abs o_k \right) \AND  \equant {o_{lsb}} F \right)
\end{align}
The last predicate has signature $(i, \abs{o}_{msb})$ and is equivalent to the predicate $\equant{o_{lsb}}F$ after the $o$ bit are replaced with their respective $\abs{o}$ bits. In other words, the output non-determinism is increased.
\end{example} % Output quantizer

The input quantizer is more complicated than output quantization.
\begin{example}[Input Quantizer]
Consider an interface $F(i, o)$ where $i \sequiv i_1 \cup i_2 \cup \ldots$ is a composite variable representing an infinite bit vector. Let $i_{msb} = i_1 \cup \ldots i_N$ be the most significant bits and $i_{lsb} = i_{N+1} \cup \ldots$ be the least significant bits. The input coarsening operator yields the following equivalent prediates.
\begin{align}
&\incoarsen(F, Q(\abs i, i)) \\
&= \ihide(i, \comp(Q(\abs i, i), F))\\
&=\equant i \left( \bigwedge_{k=1}^{N} \left( i_k == \abs i_k \right) \AND F \AND \uquant i \left(\bigwedge_{k=1}^{N} \left( i_k == \abs i_k \right)  \implies \NB{F}\right) \right)\\
&=\equant i \left( \bigwedge_{k=1}^{N} \left( i_k == \abs i_k \right) \AND F \AND \uquant i_{msb} \left(\bigwedge_{k=1}^{N} \left( i_k == \abs i_k \right)  \implies \uquant i_{lsb} \NB{F}\right) \right)\\
&=\equant i \left( \bigwedge_{k=1}^{N} \left( i_k == \abs i_k \right) \AND F  \right) \AND \uquant i_{msb} \left(\bigwedge_{k=1}^{N} \left( i_k == \abs i_k \right)  \implies \uquant i_{lsb} \NB{F}\right)\\
&=\equant i_{msb} \left( \bigwedge_{k=1}^{N} \left( i_k == \abs i_k \right) \AND \equant i_{lsb} F  \right) \AND \uquant i_{msb} \left(\bigwedge_{k=1}^{N} \left( i_k == \abs i_k \right)  \implies \uquant i_{lsb} \NB{F}\right)
%&= \equant i_{lsb} F \AND \uquant{i_{lsb}} (\NB{F})
\end{align}
The last predicate has signature $(\abs{i}_{msb}, o)$ and is equivalent to the predicate $\equant i_{lsb}F \AND \uquant{i_{lsb}}(\NB{F})$ after all $i$ bits are replaced with their respective $\abs{i}$ bits. The $\equant i_{lsb}F$ term increases output non-determinism by taking a union of outputs generated by different lower significant bit assignments. Term $\uquant{i_{lsb}}(\NB{F})$ imposes that if some input assignments can block, then all other input assignments with identical most significant bits are also blocked.
\end{example} % Input quantizer

\begin{remark} \label{rem:quantizer}
A multi-dimensional quantizer simply consists of multiple scalar quantizers. The encoding for the interval $[0,1]$ can be rescaled to an interval $[a,b]$. Additional care needs to be taken for non-powers of two, and overflow bits can be added to extend to $\mathbb{R}$. \end{remark}

\subsection{Lunar Lander} \label{sec:lunar}

We consider the lunar lander from OpenAI gym \cite{brockman2016openai}, a \texttt{python} simulation environment for reinforcement learning research.
The lander is set up within a physics simulation engine and has six continuous state dimensions and two continuous input dimensions. After some minor modifications \footnote{We removed a small additive actuation noise and the geometry of the lander was changed to be a square instead of a trapezoid. The trapezoid meant that the position $(p_x, p_y) = (0,0)$ did not align with the center of gravity.}, we identified the following discrete time dynamics 
\begin{align}
p_x^+ &= p_x + .01031 v_x \label{eqn:landerx} \\
p_y^+ &= p_y + .0225 v_y \\
v_x^+ &= v_x -.0539 u_1 \sin(\theta) + .0106 u_2 \cos(\theta)  \\
v_y^+ &= v_y + .0359 u_1 \cos(\theta) + .00707 u_2 \sin(\theta) - .0267 \\
\theta^+ &= \theta + .05 \omega \\
\omega^+ &= \omega - .05598 u_2 . \label{eqn:landeromega}
\end{align}
Control input $u_1 \in \{0\} \cup [.5, 1]$ represents a main thruster mounted on the bottom of the lander, and input $u_2 \in [-1,-.5] \cup \{0\} \cup [.5,1]$ represents a pair of side thrusters. Only one side thruster can be activated at a time and both are aligned in such a way that they apply both a torque and a linear force when activated. Both thrusters can only apply impulses with magnitude greater than $0.5$ when activated. This limits the system's ability to exert fine control over the system without resorting to high frequency chattering.

%Challenges include different spatial and time scales. For instance $x^+$'s update equation only shifts the system in a small way.
%Small time steps and different spatial scales. Specific dimensions depend on other ones quite a lot.

The continuous region of interest for our problem is $p_{x} \in [-1,1]$, $p_y \in [0,1.3]$, $v_x \in [-1,1]$, $v_y \in [-1,1]$, $\theta \in [-\frac{\pi}{5}, \frac{\pi}{5}]$ and $\omega \in [-.6,.6]$. The discretized state space consists of $\sim$137 billion states obtained from a $256 \times 256 \times 64 \times 64 \times 32 \times 16$ grid. 
Inputs are constrained to discrete sets $u_1 \in \{0, .66, .83, 1\}$ and $u_2 \in \{-.5, 0, .5\}$.

While one could construct interfaces associated with each equation \pref{eqn:landerx}-\pref{eqn:landeromega}, the positional components $p_x^+$ and $p_y^+$ would require finer grids than those above to capture a changing state over the short time horizon.
Instead of using the one-step dynamics above, we instead use a sampled system with a period of 9 time steps. We refer to the interfaces associated with the unrolled dynamics with the names shown in \Cref{table:landerbits}.
The time to construct all six interfaces was $430.5$ seconds. A reach objective with a target region $T$ is specified by
\begin{align*}
p_x \in [-.1,.1] \AND p_y \in [1.2, 1.3] \AND \theta \in [-.15,.15] \AND v_y \in [-8,.1] \label{eqn:target}
\end{align*}
which corresponds to 73 million states of the full dimensional space.
We iterate a custom reach operator 20 iterations to construct a winning set with 344.6 million states. Trajectories reaching the target are depicted in \Cref{fig:landerreach}. 
The control synthesis runtime was 4194 seconds but a small portion of that includes applying the coarsening operation as detailed later. 
We applied the aforementioned decomposed predecessor and greedy coarsening heuristic and gradually increase the complexity threshhold from $10^5$ BDD nodes until a cap of $10^6$ nodes. For the lunar lander problem, many computers would run out of memory before the game solver reaches a fixed point.
These abstraction and synthesis runtime numbers were achieved using additional techniques outlined below.

\begin{figure}
\centering
\includegraphics[width=.8\columnwidth]{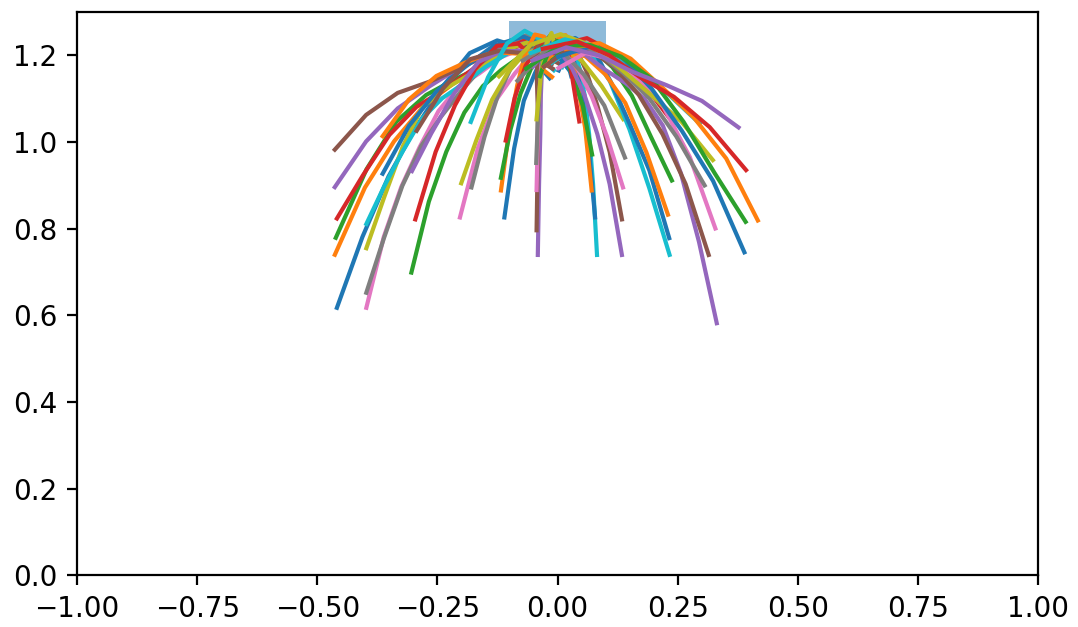}
\caption{Two dimensional positional trajectories reaching a target set in blue.} \label{fig:landerreach}
\end{figure}

\textbf{Abstraction with Heterogeneous Grids:}  While sampling over a longer time horizon lets us use a coarser grid, it comes at the cost of increasing the number of interface inputs. States $\theta$ and $\omega$ at time $t=0$ can influence state $p_x$ at future time steps through their effect on $v_x$.
While $p_x$ can depend on $\theta$ and $\omega$ over time, its sensitivity to those values is small over short horizons. Heterogeneous grids exploit this insight and allow small perturbations of $\theta$ and $\omega$ to be ignored. \Cref{table:landerbits} shows how interface $F_{p_x}$ allocates less bits to those variables than interfaces $F_{\theta}$ and $F_\omega$, which are much more sensitive to the same perturbations over the same time horizon. Curiously, $F_{p_x}$ assigns zero bits to $\omega$. This signifies that it only assumes that $\omega \in [-.6,.6]$ but does not otherwise care what the specific value is due to the low sensitivity.

\begin{table}
\centering
{
\newcommand\bits[2]{#1 \leftarrow #2}
\begin{tabular}{|c|c|c|}\hline 
Interface & Output  & Input \\
Name & Variable  & Variables \\ \hline 
$F_{p_x}$ & $\bits{p_x^+}{8}$  & $ \bits{p_x}{8}$, $\bits{v_x}{6}$, $\bits{\theta}{3}$, $\bits{\omega}{0}$, $u_1$, $u_2$ \\ \hline
$F_{p_y}$ & $\bits{p_y^+}{8}$  & $ \bits{p_y}{8}$, $\bits{v_y}{6}$, $\bits{\theta}{3}$, $\bits{\omega}{0}$, $u_1$, $u_2$ \\ \hline
$F_{v_x}$ & $\bits{v_x^+}{6}$  & $\bits{v_x}{6}$, $\bits{\theta}{5}$, $\bits{\omega}{4}$, $u_1$, $u_2$ \\ \hline
$F_{v_y}$ & $\bits{v_y^+}{6}$  & $\bits{v_y}{6}$, $\bits{\theta}{5}$, $\bits{\omega}{4}$, $u_1$, $u_2$ \\ \hline 
$F_\theta$ & $\bits{\theta^+}{5}$ & $\bits{\theta}{5}$, $\bits{\omega}{4}$, $u_2$ \\ \hline 
$F_\omega$ & $\bits{\omega^+}{4}$ & $\bits{\omega}{4}$, $u_2$ \\ \hline 
\end{tabular}
\caption{Each row represents an interface of the unrolled dynamics. The notation $\bits{p_x}{8}$ signifies that $p_x$ is a continuous variable and the interface has allocated $8$ bits to it. Different interfaces can have different views of the same variable, e.g. interface $F_\omega$ allocates $4$ bits to $\omega$ while $F_{p_x}$ allocates none. Inputs $u_1$ and $u_2$ are discrete so they do not adopt this notation.}\label{table:landerbits}
} 
\end{table}

\textbf{Abstraction with Coarse Shifted Grids:}
%Applying \Cref{alg:disjunctive} to the input space granularities provided in \Cref{table:landerbits} 
%While \Cref{alg:disjunctive} exhaustively traverses up to a half million elements interface input space with the granularities specified in \Cref{table:landerbits} can require up to a half million iterations
Some interfaces can have up to a half million inputs with the granularity provided in \Cref{table:landerbits}. Instead of exhaustively iterating over all of them, we use the overlapping property of shared refinement to reduce the number of iterations while still inducing a grid of the desired granularity. \Cref{fig:overlapping} depicts a simple two dimensional version of the iteration procedure where a higher granularity grid is constructed by two coarser grids that are offset from one another. This technique generally yields a more conservative abstraction than the one obtained by a full granularity traversal, but leads to a reduction in abstraction runtime. An overapproximation of a forward reachable set was represented as a hyper-rectangle. Simple interval analysis techniques could be applied for the one-step dynamics, and iterated over $9$ time steps. This technique is not implemented in prior tools.
\begin{figure}
\centering
\includegraphics[width=\columnwidth]{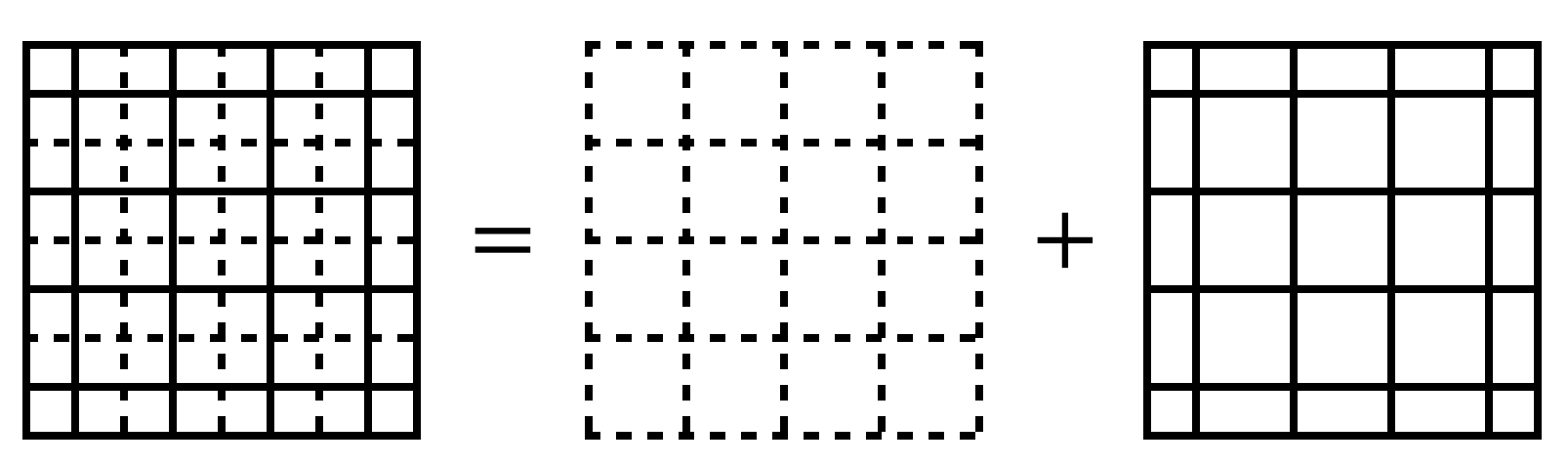}
\caption{A $8\times 8$ partition induced by overlaying a $4\times 4$ partition and a $5\times 5$ partition. Note that $4^2 + 5^2 < 8^2$. For grids of dimension $N$, this technique reduces the number of iterations by a factor of roughly $2^{N-1}$. An initial factor of roughly $2^N$ is achieved by doubling the cell width along each dimension, but iterating over two grids reduces this factor to $2^{N-1}$.} \label{fig:overlapping}
\end{figure}

\Cref{ex:sharedtodisjoint} illustrates the key idea behind how one can leverage overlapping input sets to implicitly create more discrete states than the number of loop iterations.
\begin{example}
For $j \in \{1,2\}$, let $I_j$ be a sink and $O_j$ be a source. Composing these in parallel yields two input-output interfaces $I_1 \AND O_1$ and $I_2 \AND O_2$. These sink and source interfaces represent the ones from \Cref{fig:samplequant}'s left.
The shared refinement interface outputted by $\refine(\cdot)$ has a predicate
\begin{align}
(I_1 \OR I_2) \AND (I_1 \implies O_1) \AND (I_2 \implies O_2)
\end{align}
that is logically equivalent to
\begin{align}
(I_1 \AND I_2 \AND O_1 \AND O_2) \OR (I_1 \AND \neg I_2 \AND O_1) \OR (\neg I_1 \AND I_2 \AND O_2). \label{eqn:disjshared}
\end{align}
If $I_1$ and $I_2$ correspond to disjoint sets, then this simplifies to a disjunction
\begin{align}
(I_1 \AND O_1) \OR (I_2 \AND O_2) \label{eqn:alg1out}
\end{align}
because $I_1 \AND I_2 \lequiv \false$, $I_1 \implies \neg I_2$ and $I_2 \implies \neg I_1$. Disjointness is imposed by assumption in PESSOA and SCOTS, which iterate over a partition of the space.
If $I_1$ and $I_2$ are not disjoint, then (\ref{eqn:disjshared}) can be viewed as three reach set overapproximations $O_1 \AND O_2$, $O_1$ and $O_2$ for three respective disjoint input sets $I_1 \AND I_2$, $\neg I_1 \AND I_2$, and $I_1 \AND \neg I_2$. By leveraging overlapping input domains, $\refine(\cdot)$ has generated three discrete states despite only being provided two interfaces $I_1 \AND O_1$ and $I_2 \AND O_2$.
\label{ex:sharedtodisjoint}
\end{example}

\fi % APPENDIX END 

%Even if $T$ has a compact BDD representation, continuously applying $\reach(\cdot)$ yields a sequence of BDD interfaces that grows in complexity; BDDs are implemented as graphs and their complexity is measured in the number of nodes. 

\bibliographystyle{abbrv}
\bibliography{references} 

\end{document}